\let\TeXyear\year
\documentclass{ieeeaccess} 

\let\year\TeXyear

\usepackage[hidelinks]{hyperref} 
\usepackage{orcidlink}
\definecolor{accessblue}{RGB}{0,105,154}

\usepackage{cite}
\usepackage{amsmath,amssymb,amsfonts}
\usepackage{algorithmic}
\usepackage{graphicx}
\usepackage{textcomp}
\usepackage{booktabs}

\usepackage{xcolor}
\usepackage{soul}

\renewcommand{\hl}[1]{#1}

\begin{document}
\history{}
\doi{}

\title{Two-Way Confidential VMs (2cVM): Collaborative Confidential Computing for Mutually Distrustful Parties}
\author{
\uppercase{Jordi Thijsman}~\orcidlink{0009-0007-2333-7051}\authorrefmark{1},
\uppercase{Merlijn Sebrechts}~\orcidlink{0000-0002-4093-7338}\authorrefmark{1},
\uppercase{Stefan Lefever}~\orcidlink{0000-0003-3596-1867}\authorrefmark{2},
\uppercase{Filip De Turck}~\orcidlink{0000-0003-4824-1199}\authorrefmark{1}\IEEEmembership{Fellow, IEEE},
and \uppercase{Bruno Volckaert~\orcidlink{0000-0003-0575-5894}\authorrefmark{1}.}
\IEEEmembership{Senior Member, IEEE}
}
\address[1]{Ghent University - imec, Ghent, Belgium}
\address[2]{imec, Leuven, Belgium}

\tfootnote{This research was partially supported by the imec AAA project 2cVM}

\markboth
{Thijsman \headeretal: 2cVM: Confidential Collaboration Among Distrustful Parties}
{Thijsman \headeretal: 2cVM: Confidential Collaboration Among Distrustful Parties}

\corresp{Corresponding author: Jordi Thijsman (e-mail: jordi.thijsman@ugent.be).}

\begin{abstract}
  Collaborative computation across organizations is often constrained by the
  need to process sensitive data and proprietary code without exposing them to
  untrusted infrastructure or participants. Cryptographic approaches such as
  fully homomorphic encryption and secure multi-party computation provide strong
  confidentiality but remain impractical for general workloads due to their
  extreme computational cost. We present the \textit{Two-Way Confidential
  Virtual Machine} (2cVM), a two-layer architecture that pairs a hardware
  trusted execution environment with an intra-workload isolation layer. Unlike
  regular Confidential Virtual Machines, 2cVM enforces mutual isolation between
  co-resident workloads, ensuring that participants retain control over their
  data and code. All computation in 2cVM is governed by a \textit{Commitment
  Manifest} that enumerates participants, component composition, permitted data
  channels, and authorized outputs; the manifest is locked to the VM and
  incorporated into attestation evidence, making the policy immutable and
  independently verifiable throughout the VM's lifetime. A proof-of-concept
  realization combines AMD SEV-SNP for hardware protection with the WebAssembly
  Component Model for fine-grained sandboxing of participant code. Evaluation on commodity hardware across four benchmark classes shows
  that the two isolation layers do not accumulate linearly: once a workload
  executes inside the WebAssembly sandbox, the marginal cost of enabling
  hardware memory protection is small. Overhead is workload-dependent,
  governed primarily by memory access pattern, ranging from negligible for
  sequential workloads to approximately $2\times$ for irregular,
  pointer-chasing access patterns. These results indicate that 2cVM provides a
  practical and verifiable foundation for privacy-preserving collaborative
  computation.
\end{abstract}


\titlepgskip=-15pt

\maketitle

\section{Introduction}
\label{sec:introduction}

\PARstart{M}{any} useful computations require data held by different organizations. Hospitals need to combine patient data to build reliable diagnostic models, but the data is sensitive and legally protected. Manufacturers need to combine production data with their suppliers to detect faults, but doing so exposes trade secrets. Defense systems require shared operational data to improve performance, yet that data is classified and often heavily siloed. In each case, collaboration would yield better results, but sharing the data itself carries unacceptable risk.

One of the central obstacles in inter-organizational computation is the loss of control that occurs once data leaves the environment of its owner\cite{jussen_issues_2024}. Initiatives such as European data spaces formalize data sharing agreements and governance models, but they remain largely policy- and contract-based; verifying and auditing compliance is difficult and costly. Regulatory requirements such as the GDPR add further constraints on cross-organizational data processing that any deployed system must eventually address. This challenge underscores the need for a zero-trust approach, where no infrastructure or participant is implicitly trusted, and all interactions are governed by verifiable policies.

Cryptographic approaches such as Fully Homomorphic Encryption (FHE) and secure multi-party computation (MPC) offer strong confidentiality guarantees but remain several orders of magnitude slower than native execution. Even with extensive algorithmic and hardware acceleration, recent studies show that their performance is still far from practical \cite{sidorov_comprehensive_2022}\cite{gong_practical_2024}. Consequently, these schemes are limited to specialized and latency-insensitive workloads.

This article introduces the \textit{Two-Way Confidential Virtual Machine} (2cVM), a protected execution environment for collaborative computation among mutually distrusting parties. Each participant contributes code and data as an isolated component. Interaction between components is restricted to a definition that all parties agree to in advance. These permitted interactions are recorded in a \hl{novel} \textit{Commitment Manifest}, a document, cryptographically bound to the VM's attested state, \hl{describing the future behavior of the confidential computing environment that} cannot be altered at runtime\hl{, a feature that conventional confidential computing does not offer}.

The 2cVM builds on hardware-backed confidential computing but extends its security model. A standard Confidential Virtual Machine (CVM) hides the workload from the host, but it assumes that everything inside the VM is mutually trusted. The 2cVM removes that assumption and enforces confidentiality between co-resident workloads. It provides strong, verifiable isolation at the component boundary and ensures that data only moves along authorized paths. 

This turns data-use policy into a technical property of the execution environment. Each party can verify the computational structure before contributing data, and the system \hl{offers future} guarantees that no other component can access or infer that data outside the declared channels. The result is collaborative computation in which control is retained, misuse is technically prevented, and trust between participants is no longer a prerequisite, making it suitable as a technical enforcement layer of data-sharing agreements.

The remainder of this article is structured as follows.
Section~\ref{background:main} provides the technical background on which our design builds.
Section~\ref{sec:security-model} describes the security model for the 2cVM architecture.
Section~\ref{related-work:main} examines related work both in academia and industry.
The following sections then detail our main \textbf{Contributions}:

\begin{itemize}
  \item Section~\ref{architecture:main}: the \textbf{Two-Way Confidential Virtual \hyphenation{Machine} (2cVM) architecture}\footnote{European Patent Application No. EP25187609.0 (pending, unpublished).}, which extends existing confidential computing models through a dual-layer isolation design that protects against both host infiltration and guest exfiltration. It introduces the \hl{novel abstract concept of the} \textit{Commitment Manifest} as a verifiable link between attested hardware and component-level policy enforcement \hl{throughout the VM's lifecycle}, enabling mutual assurance between data owners and code providers.

  \item Section~\ref{implementation:main}: an open-source \cite{jordithijsmanIdlabdiscover2cVMattestationagent2cVM2026a} \textbf{proof-of-concept implementation} demonstrating the practical feasibility of 2cVM on commodity SEV-SNP hardware and its integration with the WebAssembly Component Model.
  \item Section~\ref{evaluation:main}: a \textbf{quantitative validation of 2cVM's practicality} across varied categories of benchmarks, \hl{demonstrating generalizability across workloads with distinct computational profiles.}
\end{itemize}
Finally, Section \ref{limitations} discusses the limitations of the 2cVM implementation and its underlying technologies, Section \ref{conclusion:main} provides a conclusion, and Section \ref{future-work:main} outlines future research directions.

\section{Background}
\label{background:main}

\subsection{Confidential Virtual Machines}
Confidential computing (CC) is a technology that protects data in use by applying hardware-enforced encryption and integrity protections to system memory. Over the past two decades, it has evolved rapidly. Feng et al.\cite{feng_survey_2024} identify three distinct stages, the most recent beginning around 2019 with the shift from specialized, process-level enclave systems such as Intel SGX\cite{anati_innovative_2013} to general-purpose confidential virtual machines (CVM) such as AMD SEV-SNP\cite{amd_amd_2020} in 2020 and Intel TDX\cite{intel_corporation_intel_2023} in 2021. This transition expanded the protection boundary from isolated application processes to entire virtual machines, enabling full operating systems and unmodified applications to execute securely on untrusted hosts.

Remote attestation (RA) is a core mechanism in confidential computing. It allows a remote party to verify the integrity and configuration of a confidential VM before provisioning secrets or initiating execution. The hardware produces a signed attestation report containing cryptographic measurements of components that form the basis of trust, collectively referred to as the trusted computing base (TCB).

Four attestation levels (AL1 to AL4) can be distinguished for confidential VMs\cite{scopellitiUnderstandingTrustRelationships2024}. AL1 attests the physical platform and confidential hardware, that is, the processor package and on-chip security engine, and binds an attestation key to a genuine device in a known configuration. AL2 covers the platform firmware that implements memory protection and bootstraps the guest, including microcode updates, BIOS or UEFI, and the confidential runtime firmware. AL3 attests the guest operating system kernel (e.g., a Linux kernel image) that defines the protection domain inside the CVM. AL4 extends attestation to the rest of the software stack, including system services, middleware, language runtimes, and application workloads.

In practical deployments, these attestation levels are instantiated as a chain of measurements and verifications that propagates trust from the physical platform (AL1) through firmware (AL2) to the guest and, ideally, its workloads (AL3–AL4). In their most basic form, however, deployed CVM attestations focus on the lower levels, primarily AL1 and AL2, and provide assurance only about the initial launch state of the system. This is limiting for full virtual machines, which evolve as kernels, drivers, and user-space components are loaded at runtime.

Recent developments incorporate virtual Trusted Platform Modules (vTPMs)\cite{pecholtCoCoTPMTrustedPlatform2022}\cite{narayananRemoteAttestationConfidential2023a} and measured boot techniques that retain integrity evidence across attestation levels. By securely storing and extending measurements through the startup sequence, these mechanisms enable verification of a broader portion of the software stack, including the guest kernel, and selected user-space components up to AL4.

A verifier that does not want to blindly trust the cloud provider must be able to reconstruct this chain and check each link independently, which in turn requires that the components responsible for measurement and evidence generation, such as firmware, bootloaders, agents, and verification services, use open, well-specified formats and, preferably, open-source implementations. In practice, cloud providers do not always expose attestation evidence for the entire software stack, and tenants often receive guarantees about platform and launch state but little or no visibility into the code that actually executes inside the VM\cite{scopellitiUnderstandingTrustRelationships2024}. Recent policy work, including recommendations from national cybersecurity agencies, argues that open-source CVM firmware and standard attestation protocols are prerequisites for such independently verifiable, zero-trust use of confidential computing.\cite{anssi_confidential_computing_2025}

\subsection{WebAssembly}

\textbf{WebAssembly (Wasm)} is a portable binary instruction format that compiles high-level languages to a well-defined abstract machine, enabling execution across heterogeneous hardware and operating environments. Originally designed as a high-performance execution environment for browsers, its sandbox-by-default model and competitive performance have facilitated adoption in other domains such as serverless computing~\cite{kjorveziroski_webassembly_2023} and cloud-edge systems~\cite{menetrey_webassembly_2022}.

To enable execution outside the browser, the \textbf{WebAssembly System Interface (WASI)}~\cite{WebAssemblyWASI2025} was introduced. WASI defines a standardized, capability-based set of host APIs that provide controlled access to system resources such as filesystem interfaces, clocks and randomness sources. Proposals such as \textit{wasi-sockets} and \textit{wasi-http} extend this model to networking and higher-level services, enabling progressively richer application support while maintaining strict isolation boundaries. Because capabilities must be explicitly granted, modules cannot access resources they were not provided, preserving WebAssembly's sandboxed design even when performing external I/O.

To enhance modularity and interoperability among Wasm applications, the \textbf{WebAssembly Component Model}~\cite{WebAssemblyComponentmodel2025} is being developed within the WebAssembly standards community. The component model introduces a structured approach to composing applications from multiple Wasm components, describing imports and exports in the WebAssembly Interface Types (WIT) language and enabling interoperability across languages without exposing internal memory layouts. This abstraction extends portability beyond operating systems and architectures to language boundaries: for example, a Go component can invoke a Rust component through a typed interface without shared memory. Components interact only through declared functions and types, which enforces isolation and provides clear definitions of inter-component communications.

\section{Related work}
\label{related-work:main}
The following sections analyze the body of existing work, both in research and industry, and contrast it to 2cVM.
\subsection{Academic projects}
\subsubsection{Cryptographic Secure Computation}
These approaches rely on mathematical hardness to protect data in-use. Fully Homomorphic Encryption (FHE), for example allows a server to process encrypted data. They offer strong, formally provable confidentiality guarantees but have the downside that their overhead is several orders of magnitude larger than native execution. Recent studies show that FHE is still far from practical for any workloads that take longer than a few milliseconds to execute in plaintext, even with hardware acceleration~\cite{sidorov_comprehensive_2022, gong_practical_2024, marcolla_survey_2022, zhang_sok_2024}. Moreover, porting an existing algorithm to FHE often requires deep expertise in FHE and sometimes even new cryptographic primitives~\cite{viand_sok_2021}. Furthermore, standard FHE does not hide the algorithm used to process the data and does not support multiple mutually untrusting clients~\cite{gentry_fully_2009}. Circuit-private FHE~\cite{chongchitmate_circuit-private_2017} hides the algorithm from the clients, but not from the server. Although extensions such as multi-key FHE~\cite{ananth_multi-key_2020} aim to address some of these issues, they often increase the overhead even more.

\subsubsection{AccTEE}
AccTEE\cite{goltzscheAccTEEWebAssemblybasedTwoway2019} proposes a two-way sandbox for mutually distrustful workload and infrastructure providers. It combines an SGX enclave with a WebAssembly runtime and instruction-level accounting to produce an attested resource-usage log. Its trust boundary is primarily vertical: SGX protects the workload's code and data from the host, while the WASM sandbox and instrumentation protect the host and ensure an accurate representation of cloud resource usage. By contrast, 2cVM targets horizontal distrust between multiple code and data providers that share a confidential VM. Rather than metering resources for a single tenant, 2cVM uses its WebAssembly isolation layer to protect guests from one another in a collaborative computing environment.

\subsubsection{WAVEN}
WAVEN\cite{wangWAVENWebAssemblyMemory2025} targets a similar use-case to 2cVM: collaborative confidential computing with multiple distrustful parties. However, where 2cVM provides a full high-level architecture, WAVEN focuses on the technical challenge of sharing memory pages between WebAssembly modules. It replaces Wasm's linear memory with a software-managed virtual memory layer so that each module gets its own logical address space and the platform can share memory pages between modules with per-module read or read--write access. The entire runtime runs inside an SGX enclave, which hides tenant code and data from the host and lets data providers and consumers attest that the expected platform binary is executing. Its \textit{policy} is essentially which modules may map which shared pages with which permissions, inside a single fixed runtime. It does not provide an explicit, attested manifest that names participants, describes which components they contribute, or constrains which data may flow to which party. In contrast, 2cVM provides a higher-level \textit{Commitment Manifest} that is sealed into the attested state of the CVM and governs component composition and data channels; WAVEN's memory virtualization could be used to enforce such decisions inside the 2cVM runtime, but it does not address manifest-level governance by itself.

\subsubsection{Ryoan}
Ryoan~\cite{huntRyoanDistributedSandbox2017} provides a distributed sandbox for untrusted data-processing services. Its design is oriented towards a request-driven pipeline over a single user's input, defined as a directed acyclic graph (DAG) of processing services. A modified Native Client (NaCl)\cite{chrome-native-client} sandbox inside SGX enclaves runs each module, while Ryoan's runtime enforces that data flows only along the DAG edges, restricts external I/O, and resets every sandbox after each request to limit stateful leakage beyond the pipeline. This model explicitly targets single-user data-processing pipelines and does not support long-lived joint computation across data from multiple mutually distrustful parties, nor does it expose a separate, attested policy abstraction beyond the DAG structure itself. 2cVM expands beyond this per-request pipeline model: the \textit{Commitment Manifest} describes long-lived compositions of multiple datasets and code components within a confidential VM and attaches explicit permissions specifying which participants may access which inputs and outputs, with the manifest itself sealed into the VM’s attested state.

\subsection{Industry projects}
\subsubsection{Azure Confidential Clean Rooms}
\textit{Azure Confidential Clean Rooms}\cite{mathapliPerformProtectedMultiparty} aim to enable collaborative processing of sensitive data within confidential applications. However, their design remains more constrained and more dependent on external trust anchors than the 2cVM. The \textit{Clean Room Specification} defines which data sources and sinks are available, which applications may run, and which external endpoints are exposed for governance and monitoring. It also includes sandboxing and configuration metadata that roughly correspond to the scope of the 2cVM \textit{Commitment Manifest}.

Despite this resemblance, the \textit{Clean Room Specification} diverges in several important ways. It defines isolation policies rather than composition policies. Once data enters the clean room, all components appear to have mutual visibility, with no explicit mediation or selective sharing between them. Isolation is enforced only at the perimeter: policies restrict external communication but do not govern internal data flow or component interaction. In contrast, the 2cVM \textit{Commitment Manifest} captures the complete composition of components and their permitted interconnections, which allows selective and attestable enforcement of data and code boundaries inside the 2cVM.

The trust model further limits the \textit{Clean Room Specification}'s guarantees. The specification is stored externally in a \textit{Clean Room Governance Service} that relies on a distributed ledger and can be modified through consensus or administrative voting. Because of this mutability, it provides no definite guarantee about future behavior. Governance decisions, such as admitting a new application, depend on external consent authorities rather than being cryptographically bound to the execution environment. The 2cVM approach inverts this dependency by sealing the \textit{Commitment Manifest} inside the confidential environment and including it in the attested state, which gives participants cryptographic assurance that the governing policy cannot change.

Azure's implementation also faces several practical constraints. Component communication is limited to HTTP, and authorization policies are expressed only as OPA rules applied to HTTP messages. Data sharing occurs through files instead of memory or structured data exchange, which limits efficiency for multi-party workloads. Component isolation depends on containers, a weaker boundary than 2cVM's isolation mechanisms, and identity management depends on third-party external services that must be trusted.

\subsubsection{cocos.ai}
Cocos AI\cite{UltravioletrsCocos2025} provides a confidential computing platform designed for privacy-preserving machine learning workloads. Its architecture combines confidential virtual machines with in-enclave agents and encrypted communication channels across participants. A computation manifest specifies which datasets and algorithms are involved in a collaborative run, along with participant identities, and this manifest can be reflected in attestation.

Cocos is designed primarily for AI pipelines rather than broader multi-party collaboration across heterogeneous code. While it supports general-purpose workloads through containerization, this flexibility does not provide inherent isolation between workloads within the enclave, nor does it prevent workloads from colluding or attempting data exfiltration when they share the same trust domain. The computation manifest describes which components and datasets are present, but it does not define how these components compose, how data moves between them, or which internal interactions are allowed or prohibited. It therefore governs deployment rather than internal data-flow or component boundaries.

By contrast, 2cVM is designed for general collaborative computing where parties may not trust one another. The \textit{Commitment Manifest} in 2cVM specifies the component composition and the permitted data channels inside the enclave, and this specification is sealed into the attested state. This provides explicit, verifiable guarantees about how workloads may interact and how data may be accessed or released during and after execution.

\subsubsection{Enarx}
Enarx\cite{EnarxEnarx2025a} is an open-source framework designed for running unmodified applications in trusted execution environments, combining a WebAssembly runtime with attestation for Intel SGX and AMD SEV. The intent was to provide a unified approach so developers could build applications without handling hardware-specific enclave details while users benefited from stronger guarantees around data confidentiality and code integrity. To minimize the trusted computing base, Enarx runs workloads on a custom microkernel supporting a Wasm runtime.

That design choice imposed tradeoffs. The reliance on a bespoke kernel and a tightly coupled Wasmtime fork makes it difficult to adopt modern WebAssembly features such as the component model and complicates integration with multiple mutually distrustful workloads supported by 2cVM. Enarx assumes a single workload and lacks mechanisms to govern data flow or enforce mutual isolation between cooperating parties. It also depends on centralized registry and attestation services for workload distribution and trust establishment, which is a problematic model in zero-trust settings where participants avoid disclosing code or metadata to a third party.

The project has seen limited maintenance since its parent company Profian, shut down in 2023\cite{bursellClosingProfian2023}, leaving its companion services offline and documentation outdated. As a result, Enarx is impractical to deploy today.

\section{Architecture Overview}
\label{architecture:main}
This section outlines the architecture of the Two-Way Confidential VM. It first introduces the design goals, then provides an overview of the components that make up the 2cVM.
\subsection{Design Goals}
\label{architecture:design-goals}

The \textit{Two-Way Confidential Virtual Machine} architecture was designed with the following goals in mind:
\begin{enumerate}
  \item Extend the default security model of Confidential VMs with \textbf{protections against exfiltration} by malicious code, provided by one of the parties, running inside the CVM.
  \item Make this environment as \textbf{multi-purpose and generic} as possible, allowing for an arbitrary number of participants to collaboratively compute using an arbitrary number of computational components, operating on an arbitrary number of data sets.
  \item Enable \textbf{verifiable governance} in a way that is compatible, but not limited to, existing solutions such as data spaces.
\end{enumerate}

\subsection{Conceptual Components}
\label{architecture:components}

\hl{A schematic overview of the key conceptual components that make up the 2cVM} is presented in Figure~\ref{fig:arch:components}.

\Figure[h](topskip=0pt, botskip=0pt, midskip=0pt)[width=0.99\linewidth]{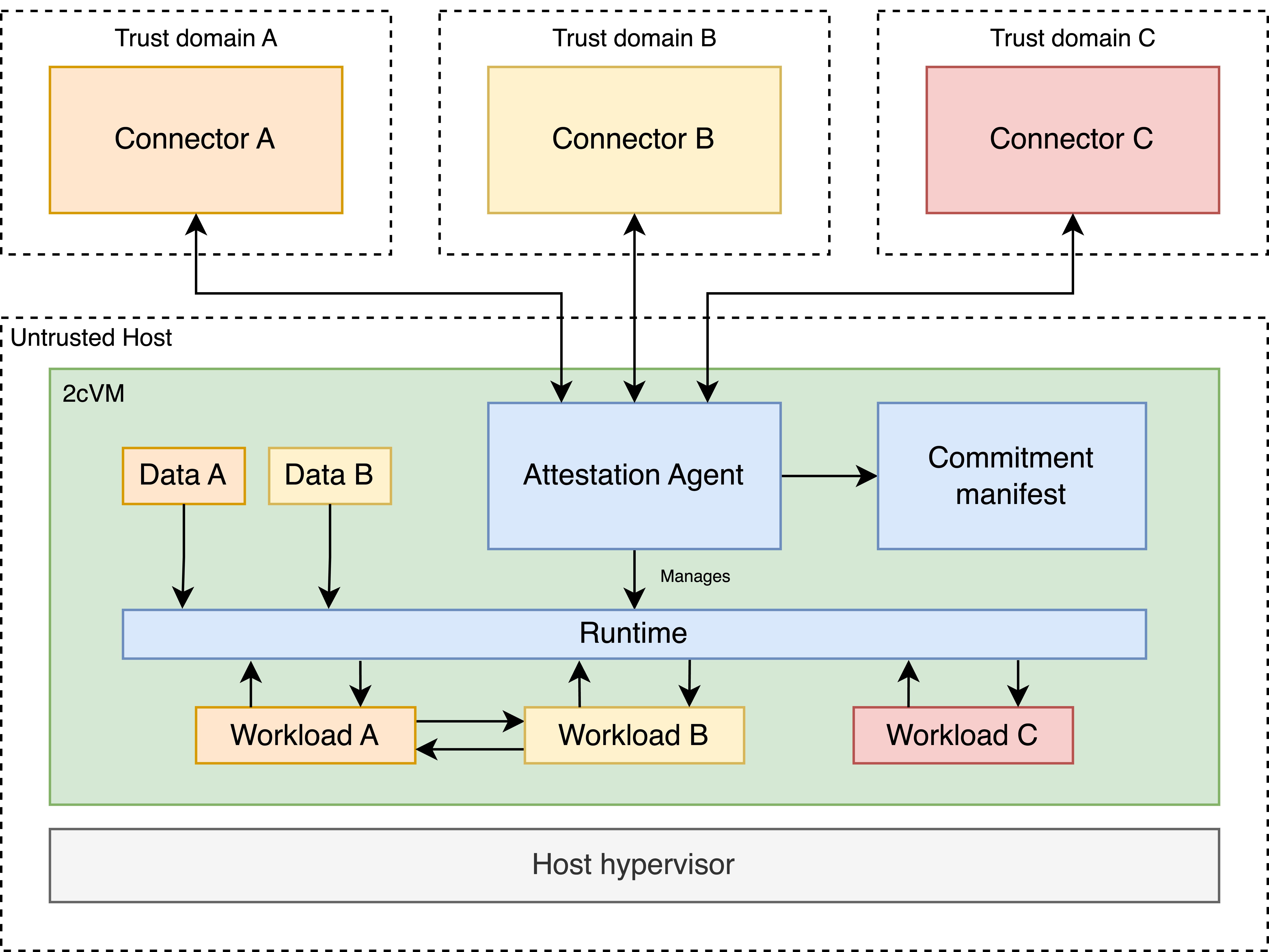}
{Overview of the \textit{Two-Way Confidential Virtual Machine} architecture.
Multiple independent trust domains contribute code and data to a shared execution
environment on an untrusted host. The \textit{Attestation Agent} enforces the
negotiated \textit{Commitment Manifest}, which governs data channels and defines the
isolation and composition of workloads.\label{fig:arch:components}}

The \textbf{Commitment Manifest} is the core directive of the 2cVM and defines all its operations. \hl{It is an abstract representation of the future behavior of the confidential computing environment. This makes it possible for the platform to not only attest its current state, but also its future behavior, even after its state and executable code changed.} \hl{On a practical level, the} manifest is a document which translates human-defined trust and security parameters into a machine-readable contract. It contains at least: a description of the participants, including an identifier used for authentication; a description of the code components, their composition, and capability-based I/O permissions; and a description of the data shared, as well as policies and ownership definitions related to this data. It is negotiated between participating parties prior to deployment.

\hl{The \textbf{Commitment Manifest Lock} is a conceptual state in which the platform code is bound to the commitment manifest in a way such that it can only perform the behavior described by the commitment manifest. As part of this state, the \textit{Commitment Manifest} is bound to the platform attestation, extending the TCB's cryptographic coverage to encompass the defined trust and security parameters at AL4. Any participant requesting attestation evidence after the commitment manifest lock goes into effect, will receive a cryptographically signed guarantee of both the current state and the future behavior of the platform.}

The \textbf{Attestation Agent} serves as the central control point for the 2cVM. It is a trusted entity in the 2cVM architecture, included in the attestation chain at AL4. \hl{It's position in the trust chain requires it to be open-source or otherwise audited by all involved parties so it can function as a guarantor of future behavior. The agent is responsible for enacting the Commitment Manifest Lock by strictly enforcing policies defined in the \textit{Commitment Manifest} such as admitting and releasing code and data, and by configuring the platform components to deny any unauthorized interference or modifications.} The \textit{Attestation Agent} provides verifiable attested evidence that both the 2cVM and the active \textit{Commitment Manifest} are in the expected trust state, allowing each party to independently confirm that the environment enforces the negotiated security constraints.

\newpage

\hl{The \textbf{Runtime} is a critical component to ensure all policies of the Commitment Manifest are fully enforced during execution of the workloads. This runtime is directed by the Attestation Agent to follow the explicitly authorized capabilities from the Commitment Manifest.} This runtime layer is not tied to a particular technology, as long as it can provide fine-grained isolation and enforce capability-bound execution. For example, using the WebAssembly Component Model, the agent can assign narrowly scoped permissions that prevent unauthorized data movement or access, creating a strict sandbox around potentially untrusted code. \hl{Important to consider is that the enforcement of the Commitment Manifest depends on the runtime's ability to isolate individual workloads. for example, the runtime is responsible for protecting against side channel attacks. More hardened solutions such as hardware-based virtualization should be considered for situations with higher security requirements.}

The \textbf{untrusted host} is a machine capable of running confidential workloads such as enclaves or confidential VMs. It provides 2cVM's second isolation layer, protecting against host infiltration. The host must be capable of attesting the state of this confidential runtime to a remote party up to and including AL2. Anything that can not be attested, such as the host kernel and hypervisor, is considered untrusted. This machine can be located either in the public cloud or the private cloud of one of the participants. A 2cVM instance is deployed to a confidential environment on this untrusted host. This instance contains all required components such as the guest kernel, guest OS, the runtime and the \textit{Attestation Agent}. This instance must support attestation up to and including AL4, which includes the \textit{Commitment Manifest}.

The \textbf{Connector} is a software component running within the trust boundary of each participating organization. Its job is to facilitate interactions with the 2cVM through the \textit{Attestation Agent}. An authorized party's connector will lock an agreed-upon \textit{Commitment Manifest} to the 2cVM. Each connector then interacts with the agent to verify the 2cVM and \textit{Commitment Manifest} attestation before sending their respective data or code into the 2cVM. Each party trusts their own connector. This connector can optionally be compatible with data spaces to provide an integration into its governance framework.

\section{Security Model}
\label{sec:security-model}
Having described the 2cVM architecture, this section outlines the security model under which the system is analyzed.

We consider a system consisting of an untrusted host platform $H$ (including hypervisor), a secure confidential computing platform $P$, and a confidential virtual machine $G$ instantiated from image $I$. The platform $P$ provides memory confidentiality, memory integrity, and remote attestation capabilities. Upon launch, $G$ undergoes a measured boot sequence (AL1--AL4), producing a measurement $\mu$ that reflects the hardware-rooted launch state, including the guest kernel and user-space runtime.

The image $I$ contains an attestation agent $A$ and a sandboxed in-guest runtime $W$, both covered by $\mu$. The runtime $W$ executes components $C_1 \ldots C_n$ over datasets $D_1 \ldots D_n$, all contributed by parties $X_1 \ldots X_n$. A commitment manifest $\sigma$ specifies the permitted component composition and allowed syntactic data access policies between components and with external I/O.

To provision data $D_n$ or code $C_n$, each party $X_n$ operates a remote connector $V_n$ within its own trusted environment. During attestation, $A$ generates an ephemeral key pair $(pk_G, sk_G)$ inside $G$.
It requests a report
\[
\rho = \mathrm{Attest}_P(\mu, H(pk_G), n),
\]
where $n$ is a nonce supplied by $V_n$ to ensure freshness.
The report $\rho$ binds the measured launch state $\mu$ to the public key $pk_G$.
The agent then signs the commitment manifest $\sigma$ using $sk_G$ and provides
$(\rho, pk_G, \mathrm{Sig}_{sk_G}(\sigma))$ to each connector $V_n$.
Each $V_n$ verifies the attestation report, checks freshness, verifies the signature
over $\sigma$, and provisions $D_n$ only if all checks succeed.

\subsection{Adversary Model}
We consider a probabilistic polynomial-time adversary that fully controls the host platform $H$, including the hypervisor, operating system, and network. The adversary may attempt to launch $G$ with a modified configuration or boot chain, replay, delay, inject, or suppress network messages, provide malicious components or datasets, and collude across multiple parties. The adversary may arbitrarily schedule, pause, or terminate $G$.

\subsection{Assumptions}
We assume that $P$ correctly enforces memory confidentiality and integrity of $G$ with respect to $H$ and produces unforgeable attestation reports. The measured launch sequence (AL1--AL4) correctly binds the guest kernel, $A$, and $W$ into the measurement $\mu$. Cryptographic primitives used for attestation and hashing are assumed secure.

\hl{As 2cVM's security posture is inherently linked to the confidential platform $P$. Security against physical attacks and side-channel leakage is thus inherited and delegated to said underlying platform $P$. The impact of this is discussed in Section} \ref{limitations:cc}.

We further assume that $W$ correctly enforces memory isolation between
components and restricts inter-component communication and access to
external resources to channels declared in $\sigma$. This enforcement is
syntactic: $W$ constrains which channels exist and which capabilities are
granted, but does not inspect or constrain the semantic content of
information conveyed through authorized channels.

\hl{
Denial-of-service attacks are considered out-of-scope. While these attacks can impact the availability of the agent and the runtime, they do not impact the security goals described below.}

\subsection{Security Goals}
The following security goals are described here and evaluated in Section \ref{sec:security-evaluation}.
\begin{enumerate}
  \item \textbf{Host Isolation.} For any adversary controlling $H$, the confidentiality and integrity of $C_n$ and $D_n$ within $G$ are preserved with respect to $H$. In particular, $H$ cannot learn plaintext contents except via outputs explicitly permitted by $\sigma$, nor modify the in-memory state of $G$, including $A$ and $W$, without detection by $P$.
  \item \textbf{Measured Launch Integrity.} Any deviation from the expected AL1--AL4 launch state, including modifications to $A$ or $W$, results in a different measurement $\mu$ and therefore an attestation report that fails verification by honest connectors.
  \item \textbf{Component Isolation and Policy Enforcement.} Assuming correct
  enforcement by $W$, data sharing between components and access to external
  resources occur only through channels declared in $\sigma$. If $\sigma$ does
  not authorize access from $C_i$ to $C_j$, then $C_j$ has no program-visible
  channel through which to observe $C_i$'s state or inputs. Similarly,
  observable outputs from any component are restricted to channels explicitly
  authorized in $\sigma$ and cannot be introduced or expanded at runtime.
  \item \textbf{Attested Composition Agreement.} All honest parties provision data only after verifying the same $\mu$ and a manifest $\sigma$ signed by a key $pk_G$ that is itself bound to $\mu$ via attestation. No party can cause execution under a different component composition or policy without detection by honest connectors.
\end{enumerate}

\section{Implementation Details}
\label{implementation:main}
This section presents the implementation of the \textit{Two-Way Confidential Virtual Machine}. The prototype demonstrates how the architecture described in Section~\ref{architecture:main} operates in practice and how its components interact within a working system. Section \ref{impl:components} discusses the implementation details of the various 2cVM components, and Section \ref{impl:flow} outlines the full workflow of a 2cVM run.

All experiments were conducted on Ubuntu~25.04 with Linux kernel~6.14, where SEV-SNP support is fully integrated into both the host and the guest. The host uses QEMU~9.2.1 with SEV-SNP enabled through standard configuration options and firmware built according to AMD's guidelines \cite{AMDESEAMDSEVSnplatest}. Within the guest, the required user-space components are installed to support attestation, composition, and execution: the \texttt{snpguest}\cite{VirteeSnpguest2025} utility for obtaining SEV-SNP attestation reports, the \texttt{WAC}\cite{BytecodeallianceWac2025} composition tool for producing composite WebAssembly binaries from the \textit{Commitment Manifest}-defined \texttt{.wac} description, the \texttt{wasmtime}\cite{BytecodeallianceWasmtime2025} runtime for executing WebAssembly components, wasm-tools\cite{wasm-tools} for checking WIT imports, and the \textit{Attestation Agent}, which coordinates the 2cVM lifecycle and invokes these tools. The agent is deployed as a \texttt{systemd} service and starts automatically when the VM boots.

To prevent data persistence beyond the VM's lifetime, all runtime operations occur within a temporary in-memory filesystem~\cite{wilke_snpguard_2024}. The directories used for component binaries (\texttt{/deps/}), input data (\texttt{/data/}), and output artifacts (\texttt{/output/}) exist only in volatile memory and are destroyed when the VM shuts down.

\subsection{Components}
\label{impl:components}
The implementation builds on two complementary isolation layers. The first layer uses AMD's SEV-SNP Confidential Virtual Machine technology as the confidential-computing substrate for protecting the 2cVM's memory and execution state from the \textbf{untrusted host}. On top of this, the WebAssembly Component Model provides a language-agnostic sandbox for isolating and composing the code contributed by each participant. Together, these layers enforce mutual confidentiality between all parties involved.

The \textbf{Attestation Agent} is implemented in Python using FastAPI and exposes a set of endpoints that govern the 2cVM lifecycle: locking the \textit{Commitment Manifest}, attesting the environment, admitting and validating artifacts, orchestrating execution, and returning results. Communication with the runtime (\texttt{wasmtime}) and attestation utility (\texttt{snpguest}) occurs via subprocess calls, which simplifies the implementation and preserves modularity. In a production deployment, these components could be linked directly to reduce process overhead and further shrink the trusted computing base.

The \textbf{Commitment Manifest} is realized as a JSON document that encodes the relations between participants, data assets, and code components in a machine-readable form. Each participant entry contains a human-readable name and a stable identifier that is used by external governance or identity systems. Components are listed with an explicit link to the participant that owns or supplies them and a list of permitted WIT interface imports. A dedicated \texttt{composition} field carries the WebAssembly Component Model composition that instantiates these components and exports the resulting interface. Data items are defined as named inputs associated with a specific participant, and a separate permissions section ties everything together: for each component, it specifies which data items may be accessed and which outputs may be produced, including the function that generates the output and the participant to whom it is delivered. Optional metadata fields, such as transfer or transaction identifiers, allow the same \textit{Commitment Manifest} to serve as a bridge to external data-space protocols.

External participants interact with the \textit{Attestation Agent} through \textbf{Connectors}. Each organization operates its own trusted Connector, which manages communication with the 2cVM on behalf of that participant. The Connectors use data-space-compatible message formats such as W3C Verifiable Credentials to maintain interoperability with existing data-space infrastructures while exchanging code, data, and configuration information.

For simplicity, the prototype omits transport-level encryption and endpoint authentication. In a production system, the agent would operate over TLS with identity-bound authentication such as OIDC. Integrating Remote Attestation--TLS (RA-TLS)\cite{knauth_integrating_2019} would allow attestation evidence to be embedded in the TLS handshake, creating verifiable, secure channels between participants. These extensions are considered out of scope for the proof-of-concept but are discussed in Section~\ref{future-work:main} on future work.

\subsection{Two-Way Confidential VM Execution Flow}
\label{impl:flow}

The 2cVM workflow proceeds in three sequential stages that collectively establish trust, admit artifacts, and execute the declared computation. Each stage is realized through a set of API endpoints exposed by the \textit{Attestation Agent}. Figures~\ref{fig:arch:flow} and \ref{fig:arch:sequence} provide, respectively, a schematic overview of these stages and a detailed sequence diagram.

\Figure[h](topskip=0pt, botskip=0pt, midskip=0pt)[width=\linewidth]{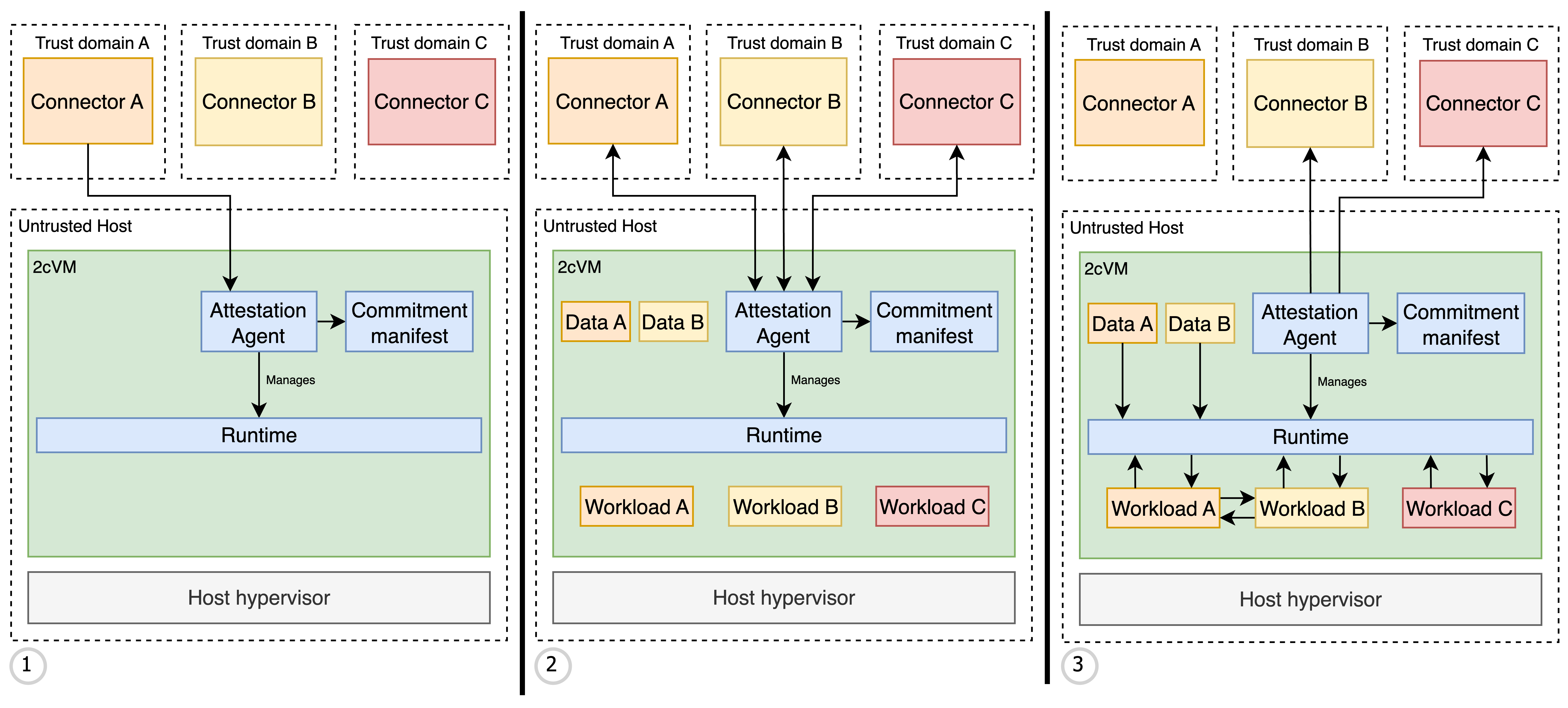}
{The different stages in the 2cVM architecture: 1) locking negotiated \textit{Commitment Manifest} to 2cVM, 2) attestation of the platform and \textit{Commitment Manifest} \& upload of sensitive information, 3) code execution and distribution of output.\label{fig:arch:flow}}

\Figure[h](topskip=0pt, botskip=0pt, midskip=0pt)[width=\linewidth]{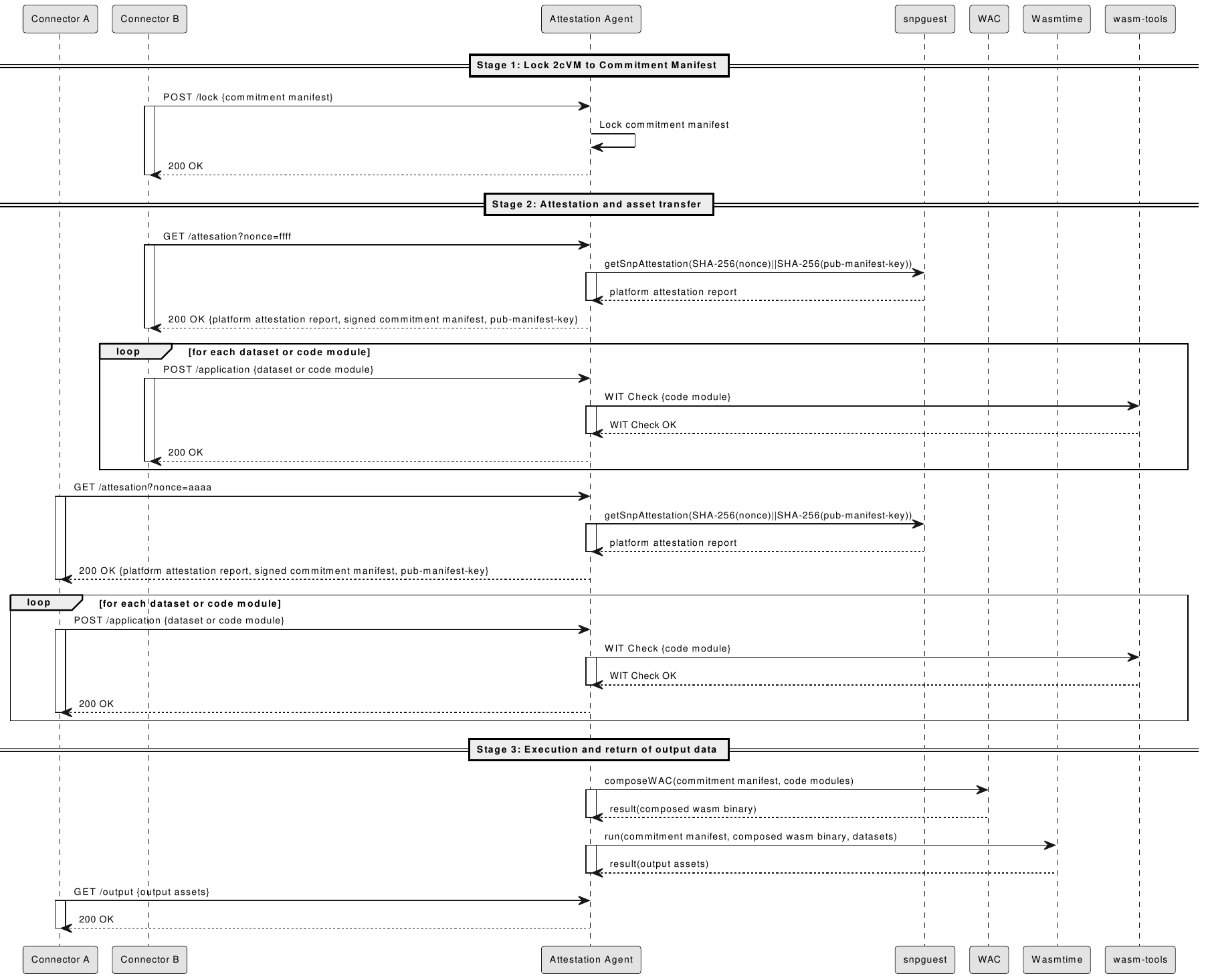}
{Sequence of a nominal 2cVM execution flow with two participants.\label{fig:arch:sequence}}
\textbf{Stage 1: Locking the \textit{Commitment Manifest}}

The first stage establishes the configuration and governance of the 2cVM instance. A participating organization begins by sending the negotiated \textit{Commitment Manifest} to the agent's \textbf{/lock} endpoint.

Upon receiving the \textit{Commitment Manifest}, the agent checks whether a lock already exists in memory. If the instance is already locked, the request is rejected with an HTTP~400 response, enforcing the manifest's immutability. Otherwise, it is serialized to the in-memory filesystem and is loaded into two thread-safe state objects: the \textit{ThreadSafeCommitmentManifest}, representing the active configuration, and the \textit{ThreadSafePartySubmissionState}, which tracks each artifact's submission status. From this point onward, the agent interprets all subsequent requests in accordance with the locked \textit{Commitment Manifest}, and the 2cVM's configuration cannot be altered without a full reset.

\textbf{Stage 2: Attestation and Artifact Submission}

Once the \textit{Commitment Manifest} has been locked, participants must verify that the system is operating on genuine confidential hardware and that the locked configuration matches the agreed-upon parameters, thereby providing proof of both 2cVM isolation layers. This verification is performed via the \textbf{/attestation} endpoint.

To prevent replay attacks, the attestation endpoint accepts a hexadecimal nonce as input. When invoked, the agent first ensures that a \textit{signing key pair} exists within the in-memory filesystem, generating one if necessary. It then constructs the attestation challenge by hashing both the provided nonce and the public part of the \textit{singing key pair}, concatenating them as
$\text{SHA-256}(\text{Nonce}) \Vert \text{SHA-256}(\text{PK}_{\text{signing}})$.
This value is passed as \texttt{user\_data} to the \texttt{snpguest} binary, which requests a hardware-signed attestation report from the CPU (AL1-AL4). The report cryptographically binds the platform state, the nonce, and the \textit{signing key pair}, extending them into the TCB.

If a \textit{Commitment Manifest} is already locked, the agent additionally signs its serialized contents with the \textit{signing key pair}, creating a \textit{commitment attestation}.

The resulting JSON response includes:
\begin{itemize}
  \item the base64-encoded SEV-SNP attestation report,
  \item the serialized and signed \textit{Commitment Manifest}, and
  \item the public part of the \textit{signing key pair}.
\end{itemize}
Participants can verify the platform attestation against AMD's endorsement keys and known measurements of the components in AL1-AL4. The \textit{Commitment Manifest} signature can be checked against the included public \textit{signing key}. This guarantees that the computation environment and policy are authentic and immutable throughout the VM's lifetime.

Once attestation has succeeded, participants may submit their declared artifacts through the \textbf{/application} endpoint. Each submission consists of a JSON structure containing a participant identifier, the artifact's manifest ID, and a base64-encoded or JSON body. In this proof-of-concept, no cryptographic verification of the identifier is performed.

Submissions are validated against the locked \textit{Commitment Manifest}.
For artifacts of type \texttt{component}, the agent decodes the binary and
places it in a quarantine area. The agent inspects the component’s declared WASI imports and compares them against the permissions declared in the Commitment Manifest for that participant. If the imports do not match the expected values, the submission is rejected and the binary is discarded. This validation step ensures that any authority to interact with external resources, such as filesystem access required to produce output artifacts, is explicitly declared through the component’s interface and auditable prior to execution. Only after
this validation succeeds is the component admitted to the in-memory filesystem
under \texttt{/deps/\textless participant\textgreater/}. For artifacts of type
\texttt{data}, the content is serialized and stored directly under
\texttt{/data/\textless participant\textgreater/}. Each successful submission
updates the \textit{Party Submission State}.

\textbf{Stage 3: Runtime Orchestration and Result Retrieval}

The third stage is triggered automatically once all submissions have been received. The agent composes the workload according to the composition defined in the \textit{Commitment Manifest}, generating a WebAssembly composition file (\texttt{.wac}) that encodes the declared component dependencies. This file is then passed to the \texttt{wac} CLI, which produces a composite WebAssembly binary that integrates all participant components within the manifest-defined structure. The resulting composite is executed using \texttt{wasmtime}, with read-write access to the \texttt{/data} and \texttt{/output} directories mounted as isolated capabilities within the runtime in accordance with the Commitment Manifest. Execution occurs asynchronously in a detached subprocess, ensuring that the agent remains responsive to API requests throughout runtime operation.

Authorized participants may later retrieve computation results through the \textbf{/application/result} endpoint. The endpoint expects a participant identifier as a query parameter, validates it against the output permissions defined in the \textit{Commitment Manifest}, and returns only the data explicitly authorized for that participant. Text outputs are returned directly, while binary outputs are base64-encoded.

This final stage enforces the end-to-end confidentiality guarantees of the 2cVM. The runtime operates strictly within the bounds of the attested \textit{Commitment Manifest}, data movement follows only the permitted channels, and results are released exclusively to authorized recipients.

\section{Evaluation}
\label{evaluation:main}
This section assesses the performance and practicality of the Two-Way Confidential VM. The evaluation is organized around four complementary benchmarks, each targeting a distinct aspect of 2cVM's behavior \hl{and collectively spanning workloads with distinct computational profiles to demonstrate generalizability across workloads.}. Section \ref{eval:FHE-2cVM} first contextualizes 2cVM against fully homomorphic encryption to establish the macro-level efficiency case. Sections \ref{eval:layer} and \ref{sec:eval:sqlite} then examine overhead layer by layer, using a compute-bound arithmetic workload and a memory-intensive SQLite workload respectively. Section \ref{sec:eval:onnx} evaluates an ONNX inference workload, exposing an important interaction between the two isolation layers under an inference workload that combines irregular
memory access with compute-intensive floating-point operations. Section \ref{sec:eval:snp-concurrency} examines the attestation subsystem under concurrent load, which is a scalability property that the other benchmarks do not exercise. Finally, Section~\ref{sec:security-evaluation} evaluates the architecture
against concrete adversarial actions within the defined security model.

\subsection{Experimental Setup}
\label{sec:setup}

All experiments were conducted on a dedicated server equipped with an AMD EPYC 8124P 16-core processor supporting SEV-SNP, 192 GB of RAM, and NVMe storage. The host runs Ubuntu 25.04 with a Linux 6.14 kernel, which provides SEV-SNP host support out of the box. Both the regular VM and the confidential VM are managed by a KVM-enabled QEMU hypervisor (v9.2.1). Each VM was provisioned with 512 MB of RAM and 2 virtual CPU cores for the arithmetic and FHE workload and 4096 MB of RAM and 1 virtual CPU core for the other workloads.
The guest disk image contains a clean installation of Ubuntu 25.04 with a Linux 6.14 kernel, providing native SEV-SNP guest support, verified using the AMD snpguest tool \cite{VirteeSnpguest2025}. The VMs were booted with custom AMD virtual firmware built according to the manufacturer's instructions \cite{AMDESEAMDSEVSnplatest}.
Each benchmark is evaluated across four configurations: native execution in a standard VM, native execution in a SEV-SNP confidential VM, Wasm component model in a standard VM, and Wasm component model in a SEV-SNP confidential VM. This structure isolates the contribution of each isolation layer independently and in combination. Benchmarks, data, and analysis scripts are provided as public artifacts \cite{jordithijsmanIdlabdiscover2cVMbenchmarks2cVMbechmarks2026}.

\subsection{Comparison with Fully Homomorphic encryption}
\label{eval:FHE-2cVM}

To contextualize 2cVM's performance profile, it is first compared against fully homomorphic encryption, which offers the strongest form of data confidentiality in use but at high computational cost. On its own, FHE protects input data against the compute provider, yet it does not provide the multi-party code and data collaboration or mutual distrust semantics that 2cVM targets. However, any practical collaborative system built around FHE or multi-key FHE would still have its runtime dominated by the underlying homomorphic evaluation. The comparison therefore focuses on the confidentiality of the compute layer using a matrix-vector multiplication workload, a fundamental linear algebra operation used throughout machine learning and data analytics. The matrix has fixed square dimensions with a matching vector length, yielding a compute-intensive task with predictable memory access patterns ideal for isolating CPU overhead. Three variants are evaluated:

\begin{enumerate}
  \item \textbf{FHE baseline.} An open-source \cite{MOZAIKSBOWinterschool20252025} matrix vector multiplication implementation which uses the OpenFHE framework \cite{OpenfheorgOpenfhepython2025} (Python interface, which internally relies on optimized C/C++ routines). This implementation performs encrypted matrix--vector multiplication using homomorphic addition and multiplication over ciphertexts, serving as a reference for cryptographic confidentiality. Each iteration of this benchmark performs the multiplication 10 times and averages the time for each operation within an iteration.
  \item \textbf{Native reference.} Implemented in Rust, compiled with \texttt{--release} optimizations using the stable toolchain (\texttt{cargo build --release}). This variant executes directly within a standard virtual machine and serves as the unencrypted baseline for both performance and correctness comparison. Each iteration of this benchmark performs the multiplication 200.000 times and averages the time for each operation within an iteration.
  \item \textbf{Component model variant.} The same Rust logic compiled to WebAssembly using the WASI-Preview2 target (\texttt{wasm32-wasip2}). This version executes within the 2cVM's WebAssembly component model runtime (wasmtime), enabling fine-grained sandboxing and capability-based I/O isolation. It contains two components: a caller and an executor. The caller generates the matrix and passes it to the executor once, who then performs the computation a number of times and eventually returns the result once. Each iteration of this benchmark performs the multiplication 200.000 times and averages the time for each operation within an iteration.
\end{enumerate}

Each variant iteration was measured ten times with random matrices. The average and sample standard deviation per configuration were computed, and the ratio of the averages was formed to obtain the relative overhead. The reported $\pm$ values are propagated standard deviations obtained by applying standard error propagation to the ratio. Matrix and vector initialization happens outside the timed execution window and is not part of the benchmark. The comparison uses a 128$\times$128 matrix multiplied by a vector of length 128.

FHE execution consistently runs about $10^5$x slower than 2cVM across our benchmark. Figure \ref{fig:eval:fhe_2cvm} illustrates this difference on a log scale. Although specialized acceleration techniques such as GPU, FPGA, or ASIC implementation can narrow this gap in certain tightly optimized workloads, they remain applicable only to specific primitives and do not eliminate FHE's fundamental performance barrier. Even with such acceleration, end-to-end applications remain several orders of magnitude slower than plaintext computation \cite{gong_practical_2024}.

\Figure[b](topskip=0pt, botskip=0pt, midskip=0pt)[width=0.99\linewidth]{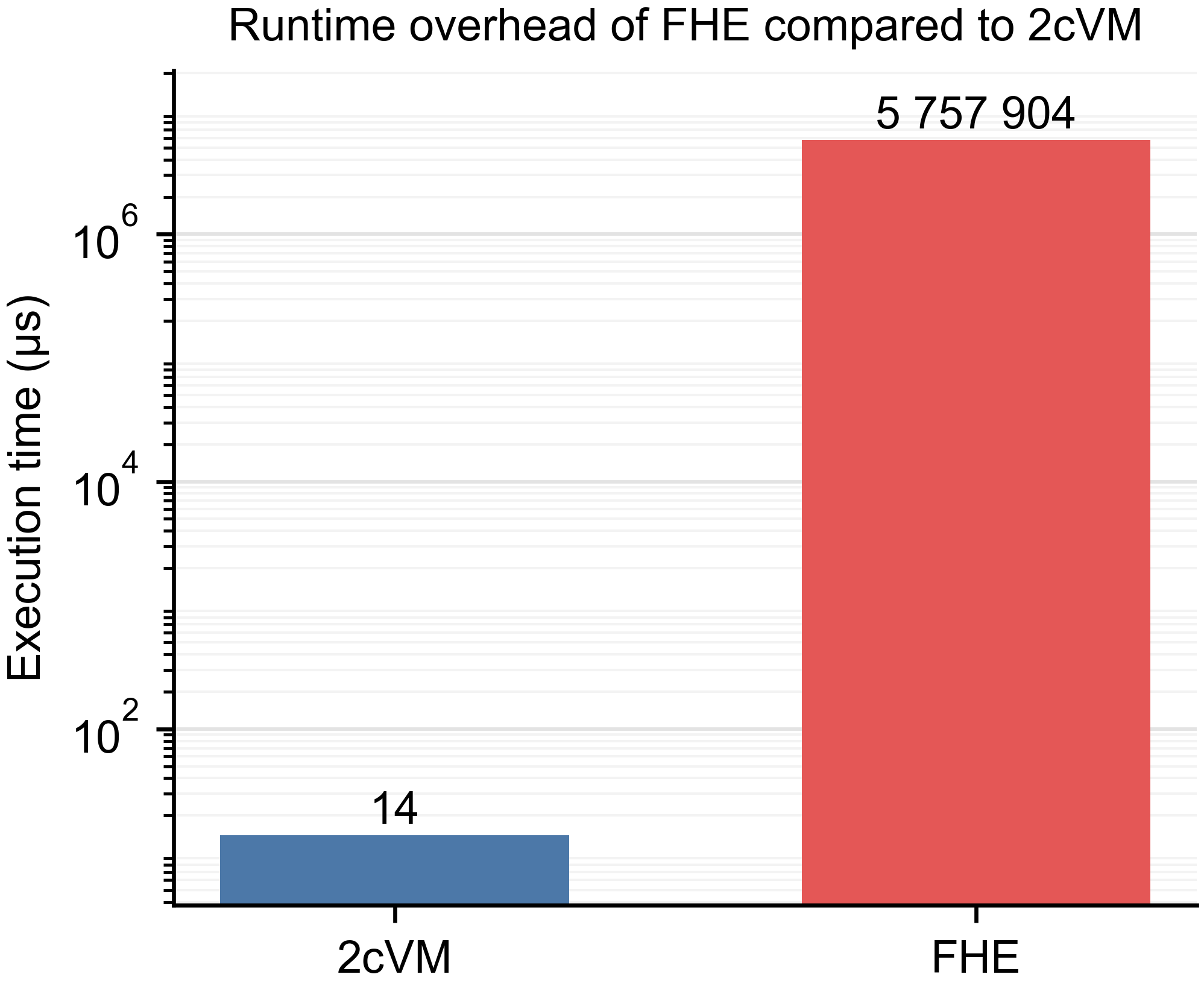}
{Comparison of 128x128 matrix--vector multiplication runtimes, showing that FHE introduces roughly a $10^5$x overhead relative to the 2cVM solution.\label{fig:eval:fhe_2cvm}}

While these findings highlight the efficiency advantage of 2cVM, the benchmark input size (128x128) is relatively small and therefore slightly biases the comparison in favor of the FHE implementation. One of 2cVM's two isolation layers, the WebAssembly component model, still incurs non-negligible overhead when transferring data across component boundaries. As matrix dimensions grow, the computational cost of the multiplication increases quadratically, whereas the data-transfer cost grows only linearly. Consequently, the relative overhead of data movement rapidly diminishes as input size increases. Figure \ref{fig:eval:input-size} visualizes this trend, showing how transfer overhead becomes negligible as input size increases. This transfer cost is a function of data volume and exchange frequency across component boundaries, not of participant count; a deployment with many participants exchanging small payloads infrequently will incur lower transfer overhead than one with few participants exchanging large data volumes repeatedly. As the WebAssembly Component Model is rapidly evolving, these data-transfer overheads should diminish over time as new and improved ways of sharing data between components are introduced.

\Figure[h](topskip=0pt, botskip=0pt, midskip=0pt)[width=0.99\linewidth]{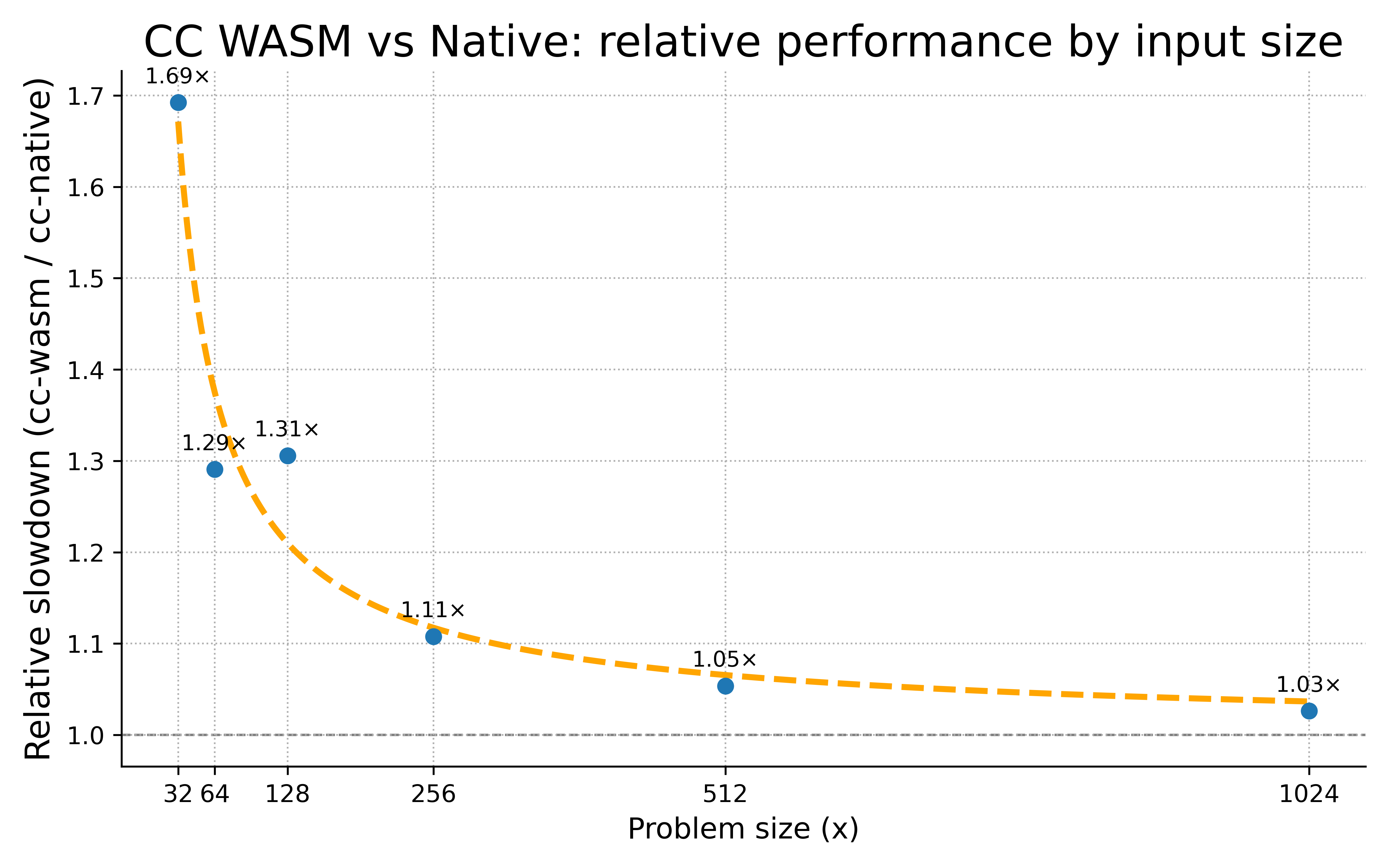}
{Impact of the problem size on the relative performance of the WASM Component Model with a fitted trend line. Overhead is dominated by data transfer between components and trends to near zero for large inputs. \label{fig:eval:input-size}}

\subsection{Layer-wise Overhead: Arithmetic Workloads}
\label{eval:layer}

2cVM employs a dual-isolation-layer strategy. The inner layer, provided by the WebAssembly Component model, prevents exfiltration by malicious code. The outer Confidential Computing layer protects the system against infiltration by the host.  This section evaluates each layer's overhead independently using the matrix-vector multiplication workload introduced in Section VII-B, which is CPU-bound, memory-resident, and syscall-light, conditions that represent a best case for 2cVM overhead. Table \ref{table:eval} provides an overview of the relative performance overhead of 2cVM and its two layers compared to a native execution; FHE is included again as a reference.

\begin{table}[h]
  \centering
  \small 
  \caption{Overhead per technology relative to native execution\\(2cVM = WASM Component + Confidential VM)}
  \resizebox{\linewidth}{!}{%
  \begin{tabular}{lll}
    \toprule
    \textbf{Technology} & \textbf{Mechanism} & {\textbf{Overhead (\%)}} \\
    \midrule
    \addlinespace[0.6em]
    2cVM & Two-layer isolation & $6.18\,(\pm\,0.20)$ \\
    \addlinespace[0.1em]
    \hspace{1em}WASM Component & \hspace{1em}WASI sandbox & \hspace{1em}$5.36\,(\pm\,0.19)$ \\
    \hspace{1em}Confidential VM & \hspace{1em}SEV-SNP & \hspace{1em}$0.77\,(\pm\,0.16)$ \\
    \addlinespace[0.6em]
    FHE & Ciphertext computation & $4.18\times10^{5}\,(\pm\,6.32\times10^{3})$ \\
    \bottomrule
  \end{tabular}%
  }
  \label{table:eval}
\end{table}

\subsubsection{\textbf{WebAssembly Component Model Layer}}
\label{eval:wasm-layer}
The inner layer of 2cVM introduces isolation of each party's data and code through the WebAssembly (WASM) component model, which enables composable modules with capability-based security. The runtime cost of this abstraction is evaluated relative to native execution.

The results show a mean overhead of 5.36($\pm$ 0.19)\% for a 512x512 matrix. As mentioned in section \ref{eval:FHE-2cVM} a good part of this overhead can be attributed to the way the Component Model handles data sharing between components and diminishes with increasing input sizes. The remaining percentage points can be attributed to the sandbox itself requiring bounds checking for memory access and allocations of new pages by the runtime. Figure \ref{fig:eval:native-wasm} shows a detailed graph of the performance overhead.

\Figure[h](topskip=0pt, botskip=0pt, midskip=0pt)[width=0.99\linewidth]{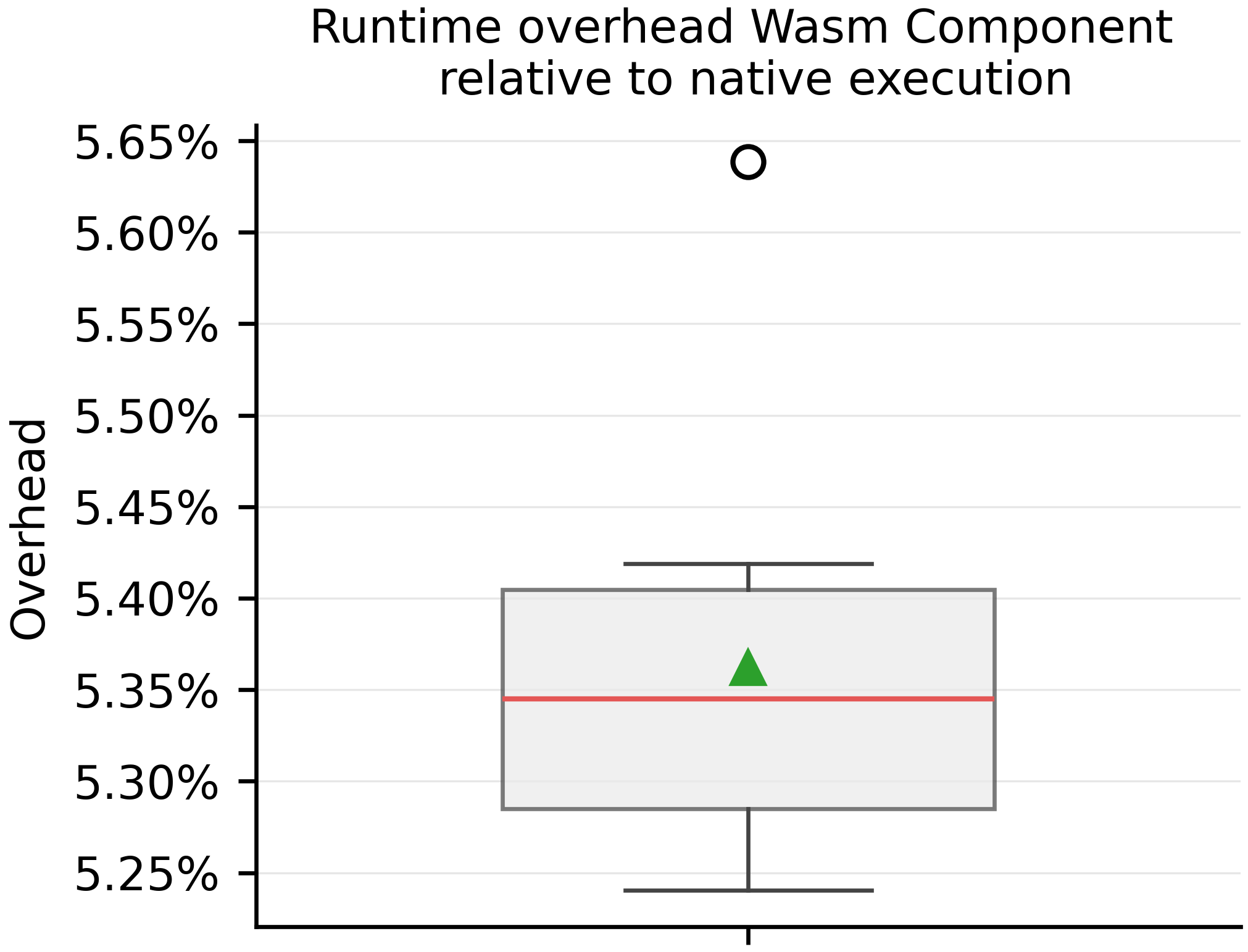}
{Impact of the WebAssembly Component isolation layer of 2cVM on execution time. Results show a mean overhead of $5.36\,(\pm\,0.19)\%$.\label{fig:eval:native-wasm}}

Despite this, the overhead is modest relative to the security and portability benefits provided by the component model. The WASM layer enforces strict module boundaries, fine-grained capability-based I/O, and language independence; all properties difficult to replicate in native code. Within 2cVM, this layer effectively compartmentalizes untrusted or third-party code while maintaining computational performance.

\subsubsection{\textbf{Confidential Computing Layer}}
\label{eval:cc-layer}
This section evaluates the overhead introduced by executing the same componentized WASM workload inside a confidential virtual machine enforced by SEV-SNP. In this configuration, both the caller and the executor components remain unchanged; the only difference lies in the underlying hardware isolation and encrypted memory model.

The results, relative to the non-confidential VM, show an average overhead of less than 1\%. Figure \ref{fig:eval:wasm-cc} shows a detailed graph. This negligible difference aligns with expectations given the characteristics of our workload: the matrix--vector multiplication benchmark is CPU-bound, memory-resident, and syscall-light. All computation occurs within user space, with minimal host--guest interaction and virtually no external I/O. Under such conditions, SEV-SNP's memory-encryption and integrity-verification engines operate nearly transparently.

\Figure[h](topskip=0pt, botskip=0pt, midskip=0pt)[width=0.99\linewidth]{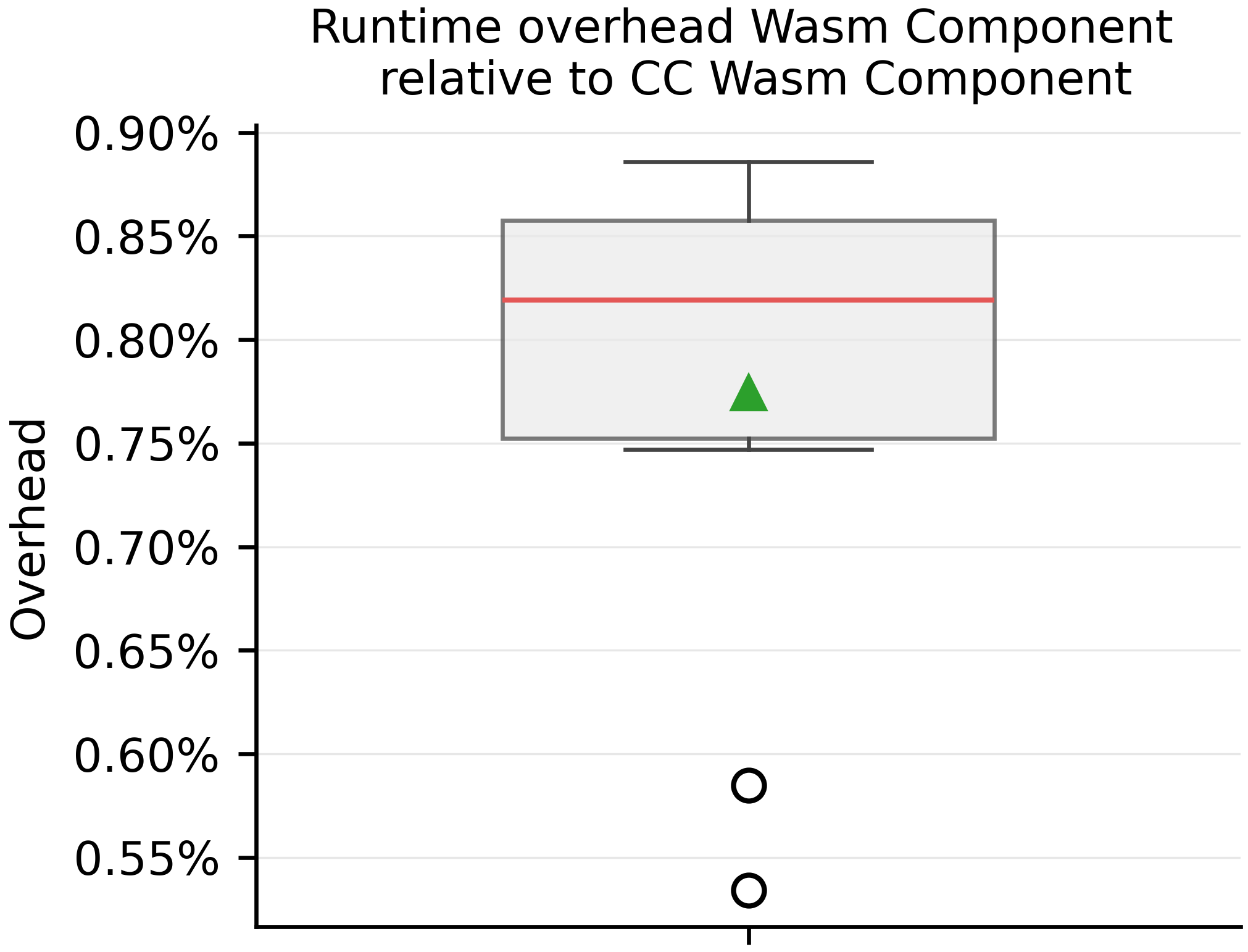}
{Impact of the confidential VM isolation layer of 2cVM on execution time. Results show a mean overhead of $0.77\,(\pm\,0.16)\%$.\label{fig:eval:wasm-cc}}

This result reflects the baseline performance cost inherent to encrypted memory access within confidential VMs. Studies that evaluated I/O-intensive or syscall-heavy workloads, such as OLTP databases, key-value stores, and filesystem benchmarks, observe an increased slowdown once execution interacts with untrusted components of the host\cite{qiuPricePrivacyPerformance2024}. The main sources of this additional overhead are repeated VM exits for emulated I/O, bounce buffering between encrypted guest memory and shared host buffers, and, on newer hardware, integrity checks on memory pages via the Reverse Map Table (RMP). These mechanisms add extra data copies, address translations, and synchronization events to each I/O operation, which amplifies latency and CPU utilization \cite{yanPerformanceOverheadsConfidential2023}. The arithmetic benchmark avoids all of these paths.

Combining layer \ref{eval:wasm-layer} and \ref{eval:cc-layer}, the holistic overhead of the Two-Way Confidential VM is demonstrated to be $6.18\,(\pm\,0.20)\%$ (Figure \ref{fig:eval:full-cvm}).

\Figure[h](topskip=0pt, botskip=0pt, midskip=0pt)[width=0.99\linewidth]{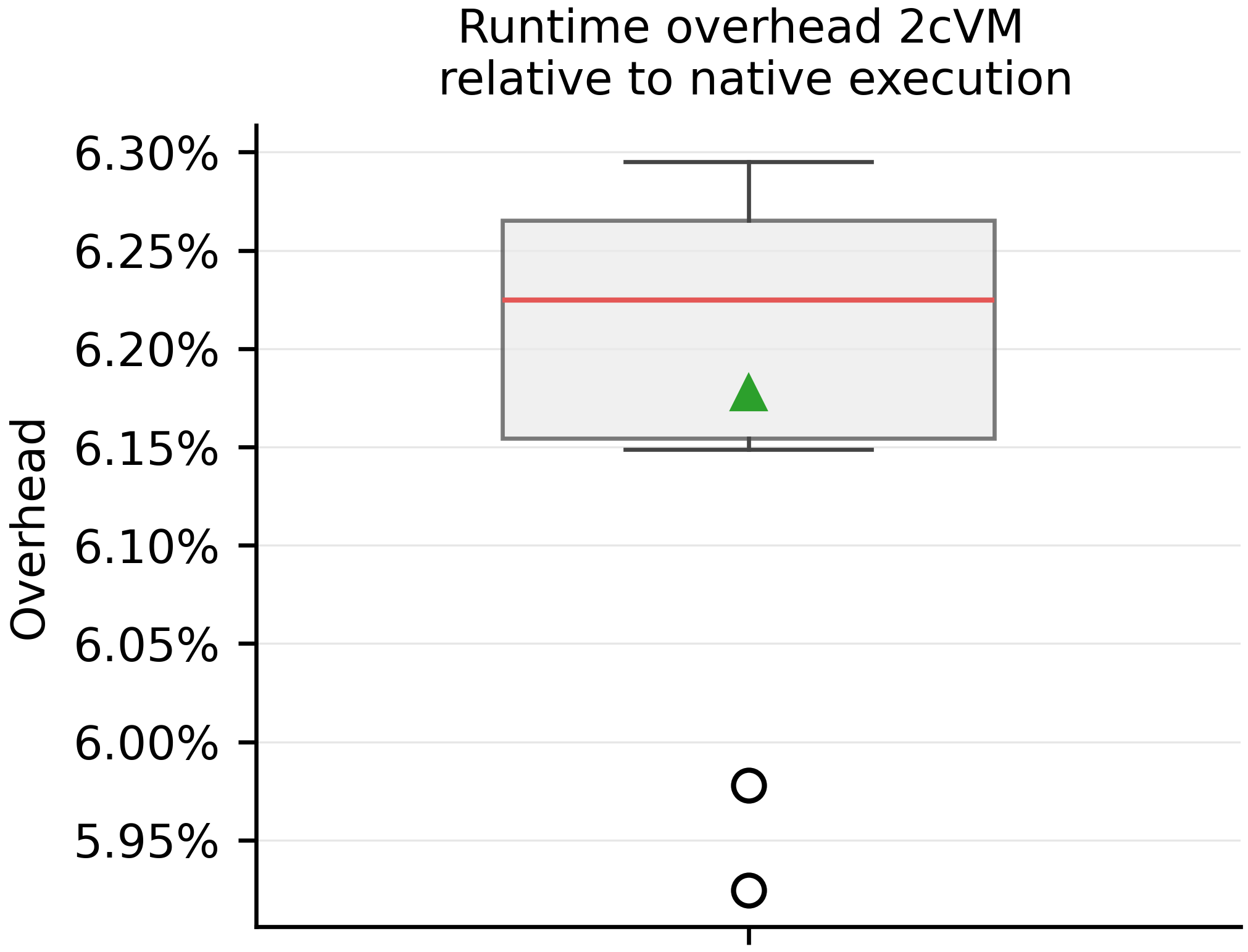}
{Impact of the full 2cVM isolation on execution time. Results show a mean overhead of $6.18\,(\pm\,0.20)\%$.\label{fig:eval:full-cvm}}

\subsection{Memory Evaluation: SQLite}
\label{sec:eval:sqlite}

The arithmetic benchmark in Section~\ref{eval:layer} isolates CPU-bound computation with predictable memory access. To evaluate 2cVM's performance on qualitatively different memory behavior, this section examines an in-memory SQLite database exercising bulk sequential writes, random indexed reads, and full table scans. These patterns differ fundamentally in their memory access behavior: bulk inserts and indexed point reads involve irregular memory access where each address depends on the result of the previous one, while full table scans access memory in a largely sequential order. The random access pattern is representative of many real collaborative workloads, such as record matching, provenance lookup, and policy evaluation, where data access is irregular and cannot benefit from hardware prefetching.

\subsubsection{Benchmark Design}

The benchmark is implemented as a Rust library exposing a single function,
\texttt{request(query: String)}, which dispatches SQL strings to a persistent
in-memory SQLite connection held in thread-local storage. The table is
declared \texttt{WITHOUT ROWID}, ensuring that every point lookup traverses the index by repeated key comparison and maximizing irregular memory access behavior.

Three sub-benchmarks are run sequentially against an in-memory database:

\begin{enumerate}
  \item \textbf{Bulk inserts.} Five million rows of the form \texttt{user\_i}
  are inserted within a single explicit transaction. Batching all inserts into
  one transaction is essential to suppress per-statement flush overhead and
  isolate runtime cost. The metric is inserts per second over the full
  transaction.

  \item \textbf{Random point reads.} Two hundred thousand indexed lookups are
  issued using the deterministic scatter pattern
  $\text{idx} = (i \times 97) \bmod 5\,000\,000$. Because 97 is coprime with $5 \times 10^6$, the sequence forms a
  permutation of the full keyspace over a complete period, ensuring the
  200\,000 samples are uniformly distributed without a random number generator.

  \item \textbf{Full table scans.} Five sequential \texttt{SELECT *} queries
  are issued over all five million rows. The result handler counts rows without deserializing column values, isolating traversal overhead from serialization cost. The metric is seconds per scan.
\end{enumerate}

Each sub-benchmark is preceded by a warm-up query to avoid penalising cold
cache effects. The same four configurations used in Section~\ref{eval:layer}
are evaluated: native execution in a non-confidential VM, native execution in a
SEV-SNP VM, Wasm component model in a non-confidential VM, and Wasm component
model in a SEV-SNP VM.

\subsubsection{Results}

Figure~\ref{fig:sql} shows throughput for each sub-benchmark, normalised to
the native non-confidential baseline.

\Figure[h](topskip=0pt, botskip=0pt, midskip=0pt)[width=\linewidth]{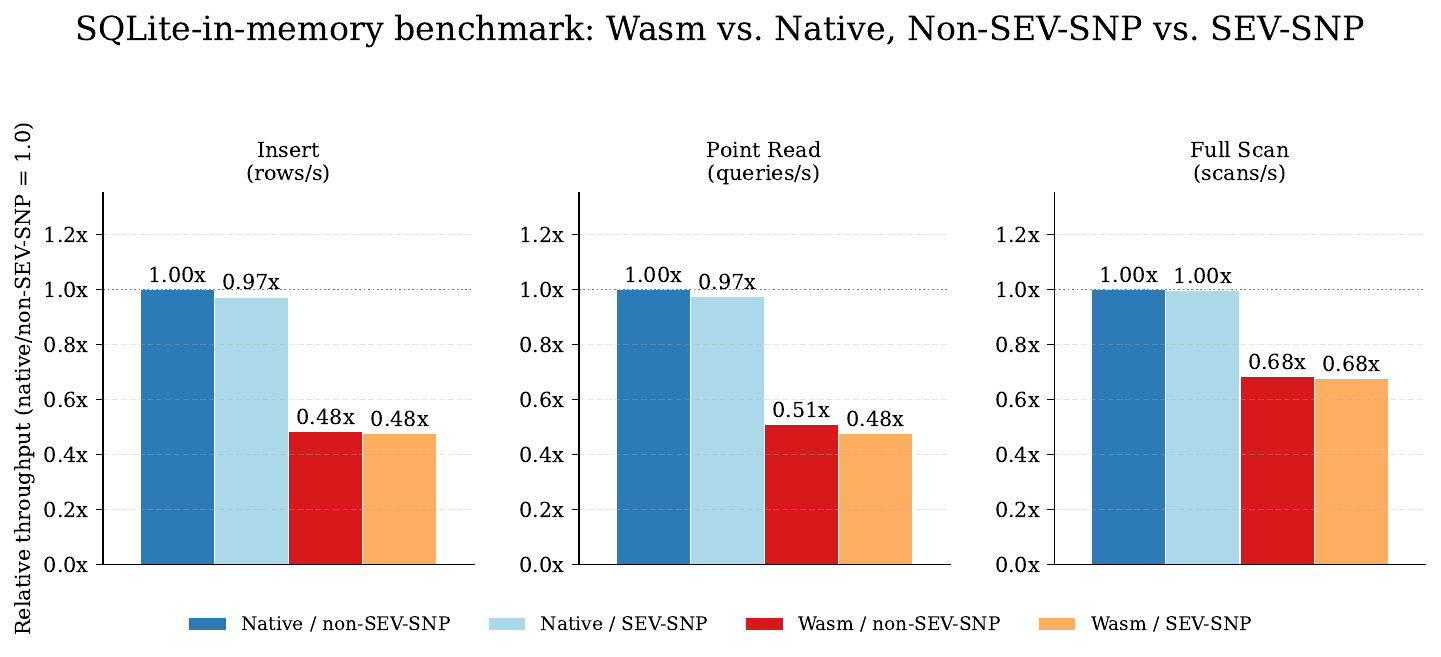}
{SQLite in-memory benchmark throughput for bulk inserts, indexed
point reads, and full table scans. Values are normalised to native execution
in a non-confidential VM. SEV-SNP introduces less than 3\% overhead across
all workloads. The Wasm execution layer imposes approximately $2\times$
overhead on write and random-access workloads and $1.5\times$ on sequential
scans.\label{fig:sql}}

The results exhibit the same two-factor structure observed in Section \ref{eval:layer}. SEV-SNP introduces approximately 3\% overhead across all three workloads. While higher than the sub-1\% observed for the arithmetic benchmark, this remains modest even for these more realistic, memory-intensive workloads, and is not the dominant source of overhead. Wasm, by contrast, imposes consistent but workload-dependent overhead: bulk inserts and point reads run at 48–51\% of native throughput (approximately 2$\times$ slowdown), while full scans run at 68\% of native throughput (approximately 1.5$\times$ slowdown). In all cases, Wasm overhead is independent of whether SEV-SNP is enabled, confirming that the two isolation layers do not interact and that the bottleneck lies entirely within the Wasm runtime.

The asymmetry between point reads (2$\times$) and full scans (1.5$\times$)
is consistent with a performance penalty that scales with the degree of
irregular memory access in the workload.

\subsubsection{Wasm Memory Overhead: Sequential vs. Irregular Access }

Two targeted memory benchmarks disambiguate the source of the Wasm overhead.

\textbf{Sequential bandwidth.} Table~\ref{tab:stream} shows STREAM \cite{McCalpin1995} results
for native and Wasm builds. Sequential memory bandwidth under Wasm is not
degraded; across all four kernels, the Wasm build matches or marginally
exceeds native throughput. This rules out memory bandwidth as the cause of
the SQL slowdown and establishes that the Wasm runtime does not impede
sequential memory operations.

\begin{table}[h]
  \centering
  \caption{STREAM memory bandwidth (best rate, MB/s), mean $\pm$ std over 5 runs.
  Wasm sequential bandwidth is within measurement noise of native.}
  \label{tab:stream}
  \begin{tabular}{lrrr}
    \toprule
    \textbf{Kernel} & \textbf{Native (MB/s)} & \textbf{Wasm (MB/s)} & \textbf{Ratio} \\
    \midrule
    Copy & $26,173 \pm 492$ & $26,835 \pm 466$ & $1.03\times$ \\
    Scale & $30,324 \pm 263$ & $31,993 \pm 564$ & $1.06\times$ \\
    Add & $30,778 \pm 173$ & $32,557 \pm 716$ & $1.06\times$ \\
    Triad & $31,062 \pm 190$ & $31,875 \pm 580$ & $1.03\times$ \\
    \bottomrule
  \end{tabular}
\end{table}

\textbf{Random-access latency.} Figure~\ref{fig:ram} shows pointer-chasing
latency as a function of working set size, measured using ram\_bench~\cite{rambench}.
For working sets fitting within L1 and L2 cache, Wasm latency is 2--3$\times$
native. In the L3 region this grows to approximately 4$\times$. Once the working
set exceeds the L3 cache, Wasm latency diverges sharply: in the DRAM regime
the overhead reaches roughly 13$\times$ native on average.

\Figure[h](topskip=0pt, botskip=0pt, midskip=0pt)[width=0.99\linewidth]{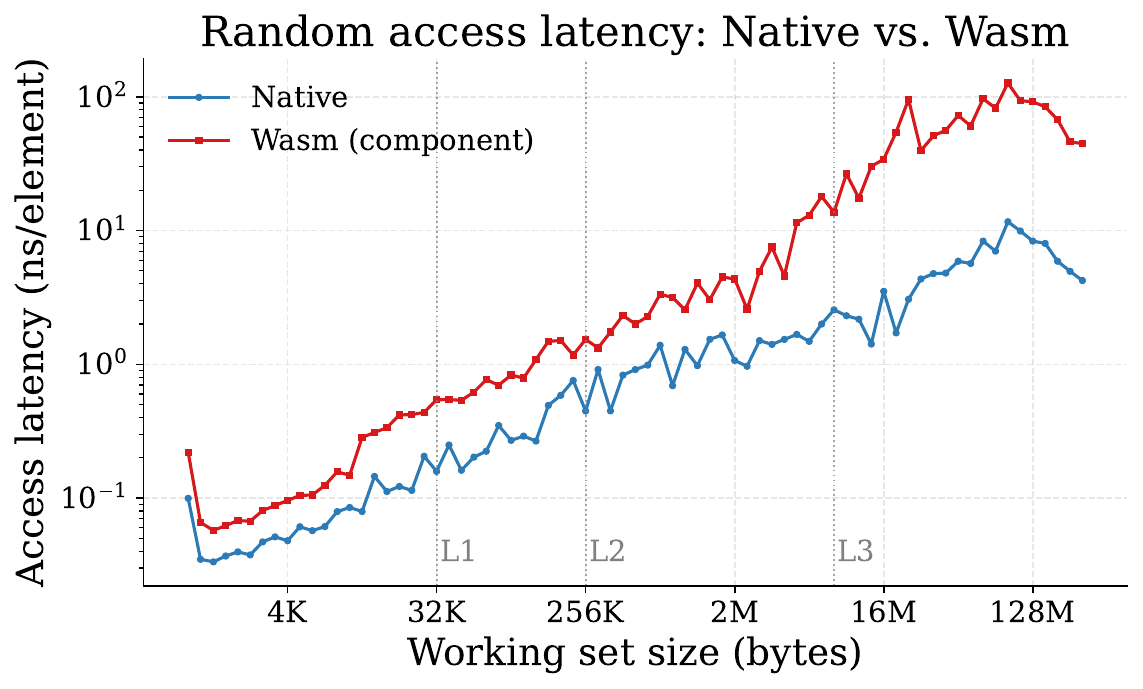}
{Random-access latency as a function of working set size for native
and Wasm component model builds. Latencies are within $2$--$3\times$ for
working sets contained in L1 or L2 cache. Beyond the L3 boundary the Wasm
curve diverges sharply, reaching around $13\times$ native latency. Vertical dashed lines mark indicate L1, L2, and L3 VM cache
boundaries.\label{fig:ram}}

These results show that 2cVM's Wasm overhead on memory-intensive workloads
is determined not by the amount of memory accessed but by the access pattern.
Workloads dominated by irregular, pointer-chasing traversal, as is common in
tree-structured data stores, graph algorithms, and sparse lookups, will
experience overhead in the range observed here, while workloads with
predominantly sequential access patterns are unaffected, as confirmed by the
STREAM results.

\subsection{Inference Workloads: ONNX MNIST}
\label{sec:eval:onnx}

The previous sections establish that SEV-SNP overhead is workload-dependent
and that Wasm overhead is driven by memory access patterns. This section
introduces a third dimension: how the two isolation layers interact under an
inference workload that combines irregular memory access with compute-intensive
floating-point operations, making it a more representative proxy for real
collaborative machine learning pipelines.

\subsubsection{Benchmark Design}

The benchmark executes ONNX-based MNIST classification compiled into three
variants:

\begin{enumerate}
  \item \textbf{Native.} A standard native binary compiled from the ONNX
  runtime, executing 5000 classification runs on a hard-coded MNIST bitmap.
  \item \textbf{Native (fast-math).} The same native binary compiled with
  aggressive floating-point optimizations enabled (e.g.,
  \texttt{-ffast-math}).
  \item \textbf{WebAssembly.} The workload compiled to WebAssembly and
  executed using the Wasmtime runtime, using ahead-of-time compiled bytecode
  (\texttt{.cwasm}) to minimize runtime compilation overhead.
\end{enumerate}

The fast-math variant is included to illustrate the performance impact of
compiler-level floating-point optimizations that are unavailable in
WebAssembly, which mandates deterministic floating-point semantics. Each
binary is executed using hyperfine \cite{Peter_hyperfine_2023}, repeated ten times, with the mean
execution time reported. Both VM configurations defined in Section~\ref{sec:setup}
are evaluated.

\subsubsection{Results}

Table~\ref{tab:onnx-results} summarizes the mean execution times. Two effects are
visible. First, SEV-SNP imposes a 20.4\% overhead on the native
binary and 16.4\% on the fast-math variant, significantly higher than observed
for the arithmetic and SQLite benchmarks. This reflects the more diverse
memory access and compute patterns of an inference workload, which stress
the hardware memory protection mechanisms more than the previous benchmarks.
Second, Wasm introduces a 12.1\% overhead relative to native in the
non-SEV-SNP environment, but SEV-SNP adds only a further 2.0\% on top. Since
the Wasm runtime already constrains execution, the marginal cost of hardware
memory protection is absorbed and the two layers do not accumulate linearly.

\begin{table}[h]
  \centering
  \begin{tabular}{lcc}
    \toprule
    \textbf{Workload} & \textbf{Native VM (s)} & \textbf{SEV-SNP VM (s)} \\
    \midrule
    ONNX (native) & $2.404\pm0.005$ & $2.894\pm0.007$  \\
    ONNX (fast-math) & $1.281\pm0.002$ & $1.492\pm0.003$\\
    ONNX (WebAssembly) & $2.696\pm0.007$ & $2.750\pm0.012$ \\
    \bottomrule
  \end{tabular}
  \caption{Mean execution time for the ONNX inference benchmark across different execution environments.}
  \label{tab:onnx-results}
\end{table}

The fast-math variant illustrates an additional tradeoff. The performance gap
between native and fast-math reflects optimizations, such as reassociation
of floating-point operations and relaxed IEEE~754 compliance, that are
fundamentally incompatible with WebAssembly's deterministic floating-point
semantics. The performance difference is therefore not overhead introduced by
the runtime, but a consequence of a deliberate design choice in the
WebAssembly specification.

\subsection{Concurrent Multi-Party Attestation}
\label{sec:eval:snp-concurrency}

Remote attestation is a core primitive in 2cVM: every participant performs exactly one attestation before submitting artifacts, making total attestation cost a direct function of participant count. Execution overhead, by contrast, is not governed by participant count but by data volume crossing component boundaries and the memory access patterns of the workloads, as the preceding benchmarks demonstrate. Scaling participant count with a fixed workload would therefore not expose meaningful variation beyond what those benchmarks already characterize. The attestation phase is where participant count has an unambiguous, direct effect, and if the underlying \texttt{/dev/sev-guest} interface serializes requests, concurrent attestation becomes the binding scalability constraint in multi-party deployments. This section empirically characterizes the latency of that interface under concurrent load.

\subsubsection{Benchmark Design}

The benchmark launches multiple worker threads within the guest VM. Each
thread opens its own file descriptor to the \texttt{/dev/sev-guest} device
and repeatedly issues the \texttt{SNP\_GET\_REPORT} ioctl to request an
attestation report. To ensure true concurrent invocation, all threads
synchronize on a barrier before each request, causing the ioctl calls to be
issued as simultaneously as possible. Each request is timed using a monotonic
clock, recording the duration of the ioctl call from invocation to completion.

For each repetition, the completion times of all concurrent requests are
ordered to determine their queue position. The first request to complete
corresponds to the effective service time of a single attestation operation,
while subsequent positions reflect the latency experienced by requests waiting
behind earlier ones.

\subsubsection{Results}

Figure~\ref{fig:latency} shows the distribution of attestation latency as
a function of queue position when ten concurrent threads request reports
simultaneously. The results exhibit a clear linear increase across queue
positions: the first request completes in approximately 4.4--4.7~ms, while
the tenth completes around 44~ms, with each successive request adding roughly
4.4~ms of additional latency. This closely matches the intrinsic service time
of a single report generation.

\Figure[h](topskip=0pt, botskip=0pt, midskip=0pt)[width=0.99\linewidth]{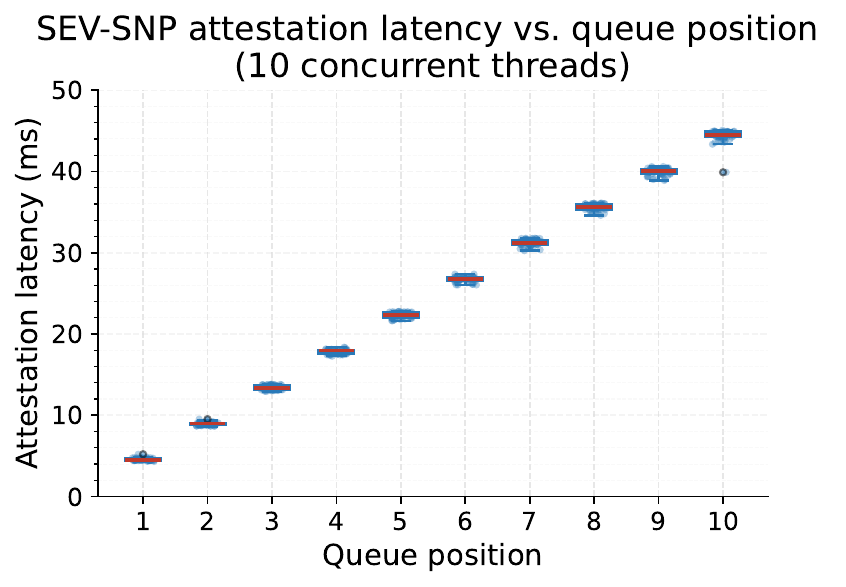}
{Attestation latency vs.\ queue position for 10 concurrent threads issuing \texttt{SNP\_GET\_REPORT} ioctls simultaneously. Latency scales linearly at $\approx$4.4\,ms per position, consistent with strict serialization at the \texttt{/dev/sev-guest} interface. Red bars indicate the median; blue points show individual observations. \label{fig:latency}}

The low variance within each queue position confirms that the dominant cost
arises from deterministic service time rather than scheduling noise or thread
contention. The \texttt{/dev/sev-guest} device effectively operates as a
single-service queue: concurrent requests are serialized, independent of how many file descriptors or threads are used. This
limitation is inherent to the attestation interface itself and is independent
of higher-level software stacks. Any 2cVM deployment with many concurrent
participants must therefore account for this linear scaling when estimating
attestation latency. Although it is not possible to parallelize this, it might be possible to instead use a vTPM for performing attestation in a more scalable manner. Future work will investigate using a vTPM for evidence attestation.

\subsection{Security Evaluation}
\label{sec:security-evaluation}

This section evaluates the 2cVM architecture against the four security
goals defined in Section~\ref{sec:security-model}, considering concrete
adversarial actions within the security model. The evaluation considers
the design of the architecture and its specific instantiation combining
AMD SEV-SNP with the WebAssembly Component Model. Limitations of the proof-of-concept implementation and the underlying
confidential computing platform are addressed in Section~\ref{limitations}.

\subsubsection{Host Isolation}

An adversary controlling the host hypervisor cannot read or modify the
in-memory state of the 2cVM instance: AMD SEV-SNP encrypts and
integrity-protects guest memory with respect to the host, and any
attempted modification is detected by the hardware. By design, all
runtime operations, including component binaries, input data, and output
artifacts, are stored exclusively in a volatile in-memory filesystem and
destroyed when the VM shuts down, ensuring that no sensitive data
persists beyond the VM's lifetime in a form accessible to the host.

\subsubsection{Measured Launch Integrity}

An adversary attempting to launch the VM with a modified kernel,
runtime, or Attestation Agent binary will produce a different
measurement $\mu$. Because $\mu$ is bound to the signing key via the
attestation report, the manifest signature returned to connectors will
be associated with a different platform state, causing connector
verification to fail. A full deployment covering AL1 through AL4
achieves this by cryptographically measuring the hardware platform,
firmware, guest kernel, runtime, and Attestation Agent via a
vTPM-based measured boot sequence. The Attestation Agent extends
this at the policy level by signing the Commitment Manifest with a key
pair generated inside the guest, extending cryptographic coverage to the
governing policy itself.

\subsubsection{Component Isolation and Policy Enforcement}

An adversary supplying a malicious component that attempts to exceed the
permissions granted in $\sigma$ is stopped before execution. Submitted
components are first placed in quarantine, where the attestation agent
inspects their declared WIT interfaces and associated WASI imports.
These imports are compared against the permissions assigned to the
component in the Commitment Manifest. If the component requests
capabilities not authorized in $\sigma$, the binary is discarded and the
submission rejected prior to execution.

If an adversarial component is admitted legitimately, it still cannot
access the internal state of other components. The WebAssembly Component
Model enforces per-component linear memory isolation by design, and
inter-component communication is restricted to explicitly declared
interfaces. Consequently, if $\sigma$ does not authorize an interface
through which $C_i$ may communicate with $C_j$, the runtime provides no
program-visible mechanism by which $C_j$ can obtain information about
$C_i$.

It is important to note that this isolation guarantee is one of
access control, not information-flow control. The Commitment
Manifest and runtime enforcement prevent a component from opening
unauthorized channels, but they impose no semantic constraints on the content
carried through authorized ones. A malicious component that is legitimately
admitted with authorized output capabilities could in principle encode and
exfiltrate sensitive information through those outputs. Preventing such covert
exfiltration would require semantic constraints on channel content, for example,
through output-filter components interposed between producers and connectors,
as discussed in Section~\ref{future-work:main}.

Finally, an adversary attempting to leak information through external
outputs is constrained by the same policy. Components may only emit data
through WASI capabilities that were authorized during admission, and
external connectors can retrieve result artifacts only if the
Commitment Manifest explicitly assigns those outputs to them.
Consequently, any observable output channel must be declared in
$\sigma$, and cannot be introduced or expanded by a malicious component
at runtime.

\subsubsection{Attested Composition Agreement}

An adversary attempting to substitute the Commitment Manifest after
locking, or to replay a stale attestation report, is defeated by two
independent mechanisms. First, the lock is immutable: any subsequent
attempt to replace the manifest is rejected by the Attestation Agent.
Second, each attestation flow requires a fresh nonce supplied by the
connector, making replay of a previous attestation report detectable.
The agent signs the locked manifest with a key pair generated inside the
guest, and the attestation report cryptographically binds the platform
measurement $\mu$ to that public key. A connector independently verifies
that the report is genuine, that the signing key is bound to $\mu$, and
that the manifest matches what was agreed. No party can cause execution
under a different component composition or policy without producing a
different $\mu$, which fails verification by honest connectors.

Finally, it is worth noting that none of the four security goals can be
violated by a crash or forced termination of the VM. An adversary able to
kill or crash the 2cVM instance can prevent computation from completing, but
the confidentiality and integrity of provisioned data and code are
unaffected: all runtime state exists exclusively in volatile memory and is
destroyed on shutdown. This reduces the impact of such events to denial of
service, which is explicitly permitted by the security model. Availability
guarantees are therefore orthogonal to the security goals defined here and
are discussed separately in Section \ref{limitations}.

\section{Discussion and Limitations}
\label{limitations}

This section discusses the limitations of the proof-of-concept implementation
and the underlying Confidential Computing and WebAssembly layers.

\subsection{2cVM Prototype}

The current implementation illustrates the technical feasibility of the 2cVM
architecture but is not intended for production deployments. It has not been
formally audited, and security guarantees are bounded by the security model
defined in Section~\ref{sec:security-model}.

\textbf{Participant authentication.} The prototype does not perform
cryptographic verification of participant identifiers on artifact submission.
Admission-time enforcement therefore relies on the confidentiality of the
communication channel rather than verified identity. A production deployment
would require identity-bound authentication, for example through OIDC, and
ideally RA-TLS to embed attestation evidence directly into the transport
layer, creating a verifiable and authenticated channel between participants
and the Attestation Agent.

\textbf{Lack of vTPM support.} The current prototype uses CPU-based evidence gathered from SEV-SNP itself. However, many operating systems implement AL4 attestation using vTPMs. On such operating systems, our solution would not correctly attest AL4, because it only uses the CPU-based evidence for attestation, instead of also including vTPM evidence. Future work should make the attestation mechanism of our attestation agent flexible so that it can choose to also include vTPM-based evidence in the remote attestation flow, for operating systems that support it.

\textbf{No recovery from mid-execution failure.} A failure during execution
produces no output and provides no recovery mechanism. In such cases the
confidential VM should be destroyed by the provider and participants should
initiate a new transaction. Data retention and cleanup policies based on
such failure scenarios are outside the scope of this work.

\textbf{Availability.} The security model does not include availability
guarantees. A malicious host can refuse to start the VM or interfere with
network access. This does not violate any of the defined security goals but
means that 2cVM provides no protection against denial of service at the
infrastructure level.

\subsection{WebAssembly}

\textbf{Single WASI interface per composition.} The WebAssembly Component
Model currently does not yet support assigning the same WASI interface to
multiple components independently within a single composition. In practice
this means only one component can hold a given WASI interface, such as
filesystem access. The agent enforces this through the Commitment Manifest,
bounding the interface to specific files or directories, but this limits
the flexibility of multi-party compositions where multiple components
require the same interface. This can be resolved using
\texttt{wasi-virt}~\cite{WASI-Virt}, which virtualizes WASI interfaces and allows
per-component isolation of shared interfaces. Moreover, this limitation is set to be removed in WASI preview 3, with its support for named interface imports. This will allow multiple distinct imports of the same interface in a single component.

\textbf{Constant-time execution.} WebAssembly currently cannot guarantee
constant-time execution, making components vulnerable to timing-based
side-channel attacks on cryptographic operations. Research on constant-time
WebAssembly exists both for compilation and
validation~\cite{9652654}\cite{gu2023constanttimewasmtimerealtime}, but
it is not yet standardized or enforced by mainstream runtimes.

\textbf{Portability of existing software.} Compiling existing software to
the WebAssembly Component Model is not always straightforward. Support is
strongest for lower-level languages such as Rust and C, while higher-level
languages have more limited or experimental support. This is an active area
of development within the WebAssembly standards community \cite{wasmcloud_language_support}.

\subsection{Confidential Computing}
\label{limitations:cc}

\textbf{Memory-only protection.} Confidential computing protects data in
use by encrypting guest memory. Data written to disk is not covered by this
protection. Care must be taken to ensure that no sensitive data is persisted
to disk during execution; the guest disk image should be integrity-protected
and, where possible, encrypted with a key held by the image provider to
prevent host interference.

\textbf{Revocation after data release.} Once output data has been released
to an authorized participant, revocation is not possible. Data release only
occurs after successful attestation, but if the attestation state were to
change after release, for example due to a firmware update altering the
measurement, previously released data cannot be recalled. This is a
general limitation of attestation-based systems and remains an open research
problem.

\textbf{Hardware vulnerabilities.} Like all confidential computing platforms,
AMD SEV-SNP is subject to hardware-level vulnerabilities. Two recent attacks
are worth highlighting for their impact on 2cVM. TEE.fail~\cite{tee-fail}
enables ciphertext side-channel attacks that can leak private keys during
cryptographic signing operations via a DDR5 memory interposer. AMD's
attestation signing happens in a separate chip and does not collapse, but
any cryptographic operation inside the guest potentially leaks a private key.
BatteringRAM~\cite{batteringramsp26} allows replay of memory pages via a
DDR4 memory interposer, enabling an attacker to bypass AMD's attestation
measurements by replaying another confidential VM's state. Both attacks
require the host owner to be actively malicious at an organizational level, a compromised host system alone is not sufficient. For this reason,
manufacturers consider these attacks out of scope, though they remain
actively exploitable.

While the prototype uses AMD SEV-SNP, the 2cVM architecture is designed to
be agnostic to the underlying confidential computing substrate. Migrating to
a more secure platform as one becomes available requires no architectural
changes. Until hardware vulnerabilities in current platforms are resolved,
deployments involving cryptographic operations inside the guest should
carefully consider the implications of these attacks within their specific
adversary model.

\section{Conclusion}
\label{conclusion:main}
This work presented the concept, design, and evaluation of the Two-Way
Confidential Virtual Machine, a system that combines hardware-backed isolation
with language-level sandboxing to protect both data and code in collaborative
computing environments. By extending the guarantees of traditional Confidential
VMs with an inner WebAssembly isolation layer, 2cVM prevents not only
infiltration from a malicious host but also exfiltration by untrusted code
running within the same confidential environment. All computation is governed
by a Commitment Manifest that is cryptographically bound to the attested
platform state, making the governing policy immutable and independently
verifiable throughout the VM's lifetime.

Experimental evaluation across four benchmark classes shows that the layered
design achieves strong security without multiplicative performance penalties.
The two isolation layers do not accumulate linearly: once a workload executes
inside the Wasm sandbox, the marginal cost of enabling SEV-SNP hardware
protection is small. Overhead is workload-dependent rather than fixed:
SEV-SNP remains modest across all tested workloads, while Wasm overhead is
governed primarily by memory access pattern, ranging from negligible for
sequential workloads to approximately $2\times$ for irregular, pointer-chasing
access patterns. The attestation subsystem introduces an orthogonal scalability
constraint, with request latency scaling linearly under concurrent load due to
serialization at the hardware interface level.

Overall, 2cVM demonstrates that hardware attestation and modular,
capability-based sandboxing can be combined to enable secure, auditable, and
efficient multi-party computation. It operationalizes zero-trust principles
within confidential computing environments, ensuring that trust is never
assumed but continuously established through attestation and manifest-based
governance, while remaining compatible with existing data-sharing frameworks
such as data spaces.

\section{Future Work}
\label{future-work:main}

\textbf{Output-filter components.}
2cVM controls the ability of components to produce external output through WebAssembly’s capability-based security model.
A component can only access system resources that are explicitly declared in its WIT interface imports. In particular,
writing output requires importing the appropriate WASI capabilities (e.g., filesystem access). As a result, any component
that is technically capable of emitting data is identifiable from its interface definition and is explicitly referenced in
the \textit{Commitment Manifest}. This provides an auditable record of which components are granted the authority to write
results during execution.

Access to those results by external parties is controlled separately. Connectors are not components within the execution
environment but external actors whose access rights are defined in the \textit{Commitment Manifest}. The manifest specifies
which connectors are allowed to retrieve particular output files produced during the computation. Consequently, both the
ability to generate outputs and the entities permitted to consume them are explicitly declared and reviewable during the
commitment phase.

While this mechanism constrains which components can produce outputs and which connectors may access them, it does not
reason about the semantic content of the produced results. A component that is legitimately granted an output capability
could still encode sensitive information in its output. Such behavior does not violate the 2cVM security model, since the
existence of the output channel and the parties allowed to receive it are explicitly declared, but it may still be
undesirable in certain collaborative scenarios.

A potential extension is the introduction of \textit{output-filter} WebAssembly components placed between producing
components and the output directory exposed to connectors. Such filters could be supplied by data providers and perform
rule-based validation, privacy sanitization, or statistical leakage checks before data becomes accessible to connectors.
Because these filters would themselves be components referenced in the \textit{Commitment Manifest} and attested as part
of the execution chain, they could provide an additional verifiable control layer over the content of released outputs.

\textbf{RA-TLS integration.} Another long-term goal is the integration of 2cVM with RA-TLS to allow remote parties to authenticate both the platform and its confidentiality commitments over standard HTTPS connections, with increased protection against man-in-the-middle attacks. This would require extending current server stacks to regenerate attestation-bound TLS certificates for each connection in order to prevent replay attacks. This regeneration is not supported in most web infrastructure today. Achieving this would unify attestation, policy enforcement, and secure data exchange in a single verifiable channel.

\textbf{Alternative inner isolation mechanisms.} The current 2cVM design relies on the WebAssembly Component Model to isolate untrusted code within the confidential VM. This model enforces strict boundaries between components and limits resource access to explicitly declared imports, providing strong, analyzable isolation. However, the programming model can be restrictive, and compiling existing software stacks to WebAssembly components is not always straightforward. In addition, the evaluation in Section \ref{evaluation:main} shows that the WebAssembly execution layer introduces measurable runtime overhead compared to native execution for certain workload classes, particularly those dominated by floating-point operations or irregular, memory-access–intensive patterns. Future work could therefore explore alternative inner isolation mechanisms that offer comparable security properties with simpler integration paths or different performance trade-offs. Potential candidates include user-space kernel interposition layers such as gVisor, which virtualize system interfaces to confine untrusted workloads. The objective is to preserve fine-grained isolation and explicit resource control while making it easier to adapt existing applications to the 2cVM architecture.

\textbf{Practical deployment considerations}. Moving from the current proof-of-concept to a production-ready system requires addressing several concerns that are orthogonal to the core isolation architecture but essential for real-world adoption. On the regulatory side, deployments that process personal data must comply with frameworks such as the GDPR, including requirements around lawful basis, data minimization, and data subject rights; the Commitment Manifest's explicit enumeration of data flows and authorized outputs provides a foundation for demonstrating compliance, but integration with organizational data protection processes remains future work. Dispute resolution mechanisms, such as tamper-evident logging of manifest negotiation and execution outcomes, are needed to support auditability when participants disagree on results or adherence to agreed terms. Operational tooling for lifecycle management, monitoring, and diagnostics of 2cVM instances is not yet developed. Finally, a cost analysis comparing 2cVM deployments against alternative approaches would inform adoption decisions but depends on workload characteristics and cloud pricing models that are outside the scope of this work.

\hl{\textbf{Formal analysis of complex threat vectors}. Future work should include a detailed formal analysis of the complete 2cVM threat model. Complex attacks such as timing attacks or other side-channel attacks should be thoroughly investigated and formally mitigated. Given multiple parties can provide untrusted code processing sensitive data, this introduces novel attack vectors making defending against timing attacks very difficult. The untrusted code could, for example, use processing delays to transmit data out of band. Formal analysis of the added attack vectors and investigation into mitigations against timing attacks is important future work. Additionally, while attacks against the WebAssembly and Confidential Computing substrates are described in literature} \cite{feng_2024}\cite{misono2024}\cite{PERRONE2025100728}\hl{, the combination of both could introduce additional attack vectors and requires a separate, hybrid analysis.}

%
%
%
\bibliographystyle{splncs04}
\bibliography{references}

\begin{IEEEbiography}[{\includegraphics[width=1in,height=1.25in,clip,keepaspectratio]{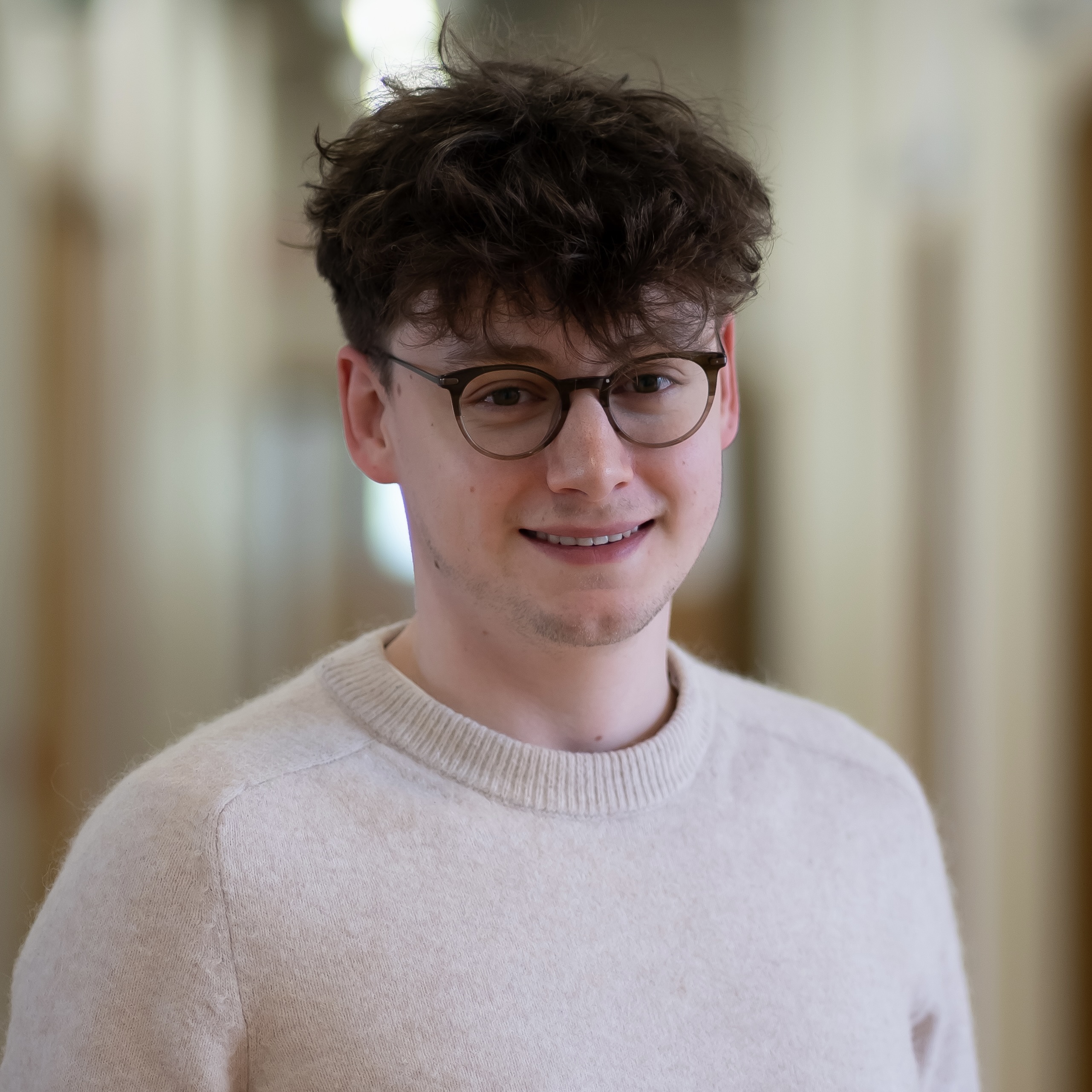}}]{Ing. Jordi Thijsman} received their B.S. and M.S. in Information Engineering Technology from Ghent University in 2023. After his M.S., he joined the IDLab research group at Ghent University - imec to pursue a PhD degree. His research focuses on provable trust and attestation across the cloud-edge continuum.
\end{IEEEbiography}

\begin{IEEEbiography}[{\includegraphics[width=1in,height=1.25in,clip,keepaspectratio]{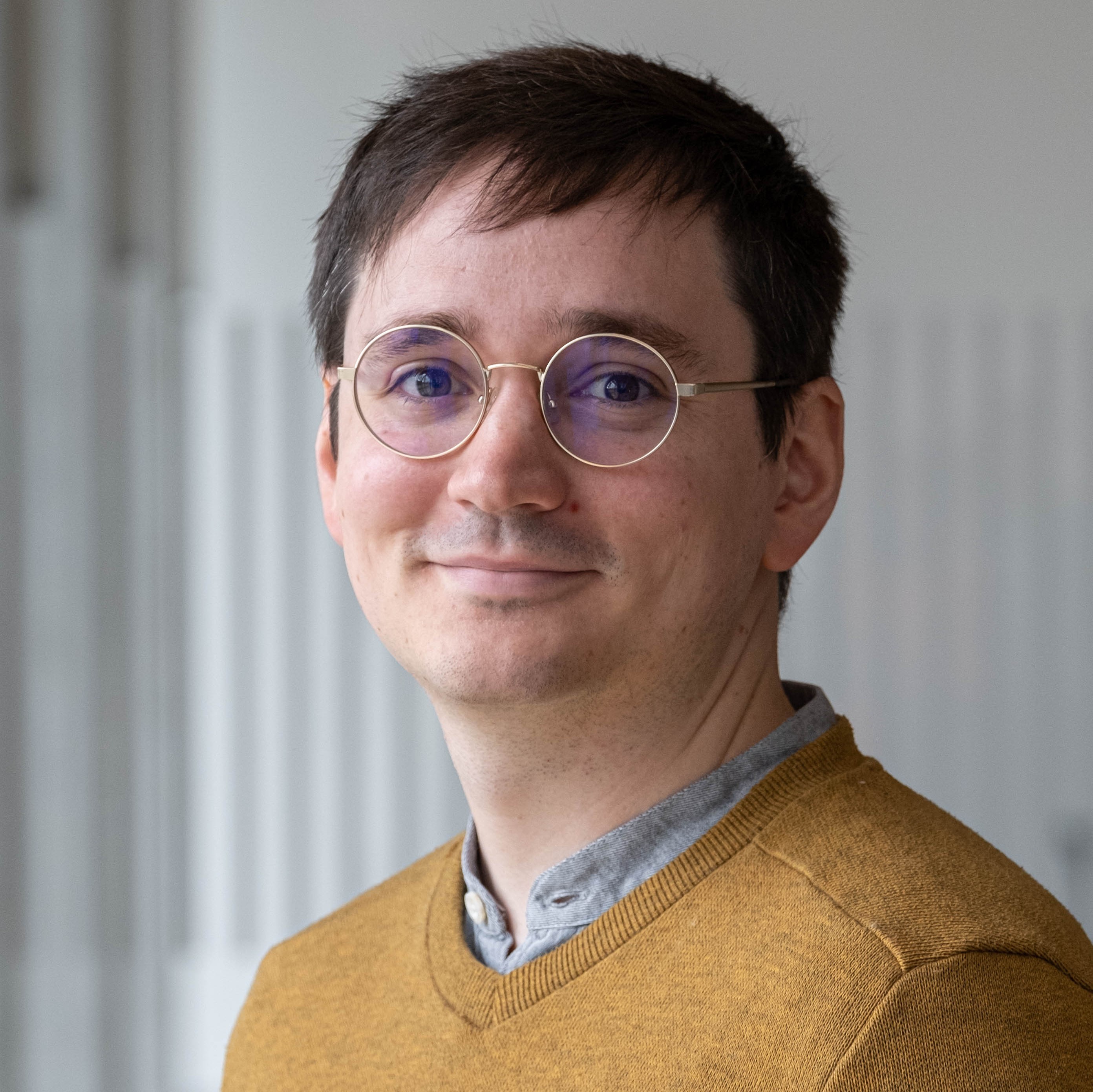}}]{Dr. ing. Merlijn Sebrechts} is a senior researcher at imec and teaches at Ghent University in Belgium. He leads a number of research tracks focused on secure and robust systems in the cloud and on devices. He is currently serving on the Ubuntu Community Council and is standardizing WebAssembly System Interfaces for IoT devices as part of the W3C and the Bytecode Alliance. He teaches topics such as Distributed Systems Design, open-source ecosystems and Computer Security. His work has been published in over 20 scientific publications and has received four awards.
\end{IEEEbiography}

\begin{IEEEbiography}[{\includegraphics[width=1in,height=1.25in,clip,keepaspectratio]{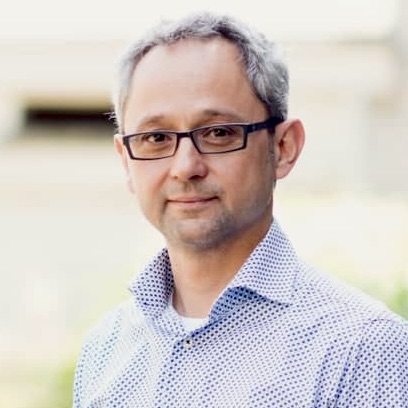}}]{Stefan Lefever} holds master’s degrees in electronics and computer science and a postgraduate degree in business administration. He has over two decades of experience in the hw-sw development in the telecom industry, where he served in roles such as R\&D engineer, system architect, and program director, developing carrier-grade telecom infrastructure and driving innovation toward SDN and NFV technologies.  Stefan joined imec to focus on secure connectivity and trusted data- and algorithm sharing architectures across societal domains. His work emphasizes the integration of advanced sensor technologies, secure IoT platforms, privacy-preserving data exchange (in data spaces) and system-of-systems inspired digital twins. Within imec’s research programs on environment, mobility, logistics, and public health, Stefan has led efforts to design architectures that combine confidentiality, integrity, and interoperability of data ecosystems and federated infrastructures. His expertise bridges research and market adoption, ensuring that secure hardware and trusted data-sharing frameworks form the foundation for scalable, real-world solutions ready to power AI solutions.
\end{IEEEbiography}

\begin{IEEEbiography}[{\includegraphics[width=1in,height=1.25in,clip,keepaspectratio]{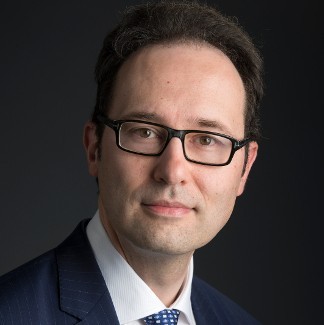}}]{Prof. Filip De Turck} (Fellow, IEEE) leads the Network and Service Management Research Group, Ghent University, Belgium, and imec. He has coauthored over 700 peer-reviewed articles. He is involved in several research projects with industry and academia. His research interests include design of secure and efficient softwarized network and cloud systems. He was elevated as an IEEE Fellow for outstanding technical contributions. He served as the Chair of the IEEE Technical Committee on Network Operations and Management (CNOM) and a Steering Committee Member for the IFIP/IEEE IM, IEEE/IFIP NOMS, IEEE/IFIP CNSM, and IEEE NetSoft conferences. He served as Editor-in-Chief of the IEEE Transactions on Network and Service Management.
\end{IEEEbiography}

\begin{IEEEbiography}[{\includegraphics[width=1in,height=1.25in,clip,keepaspectratio]{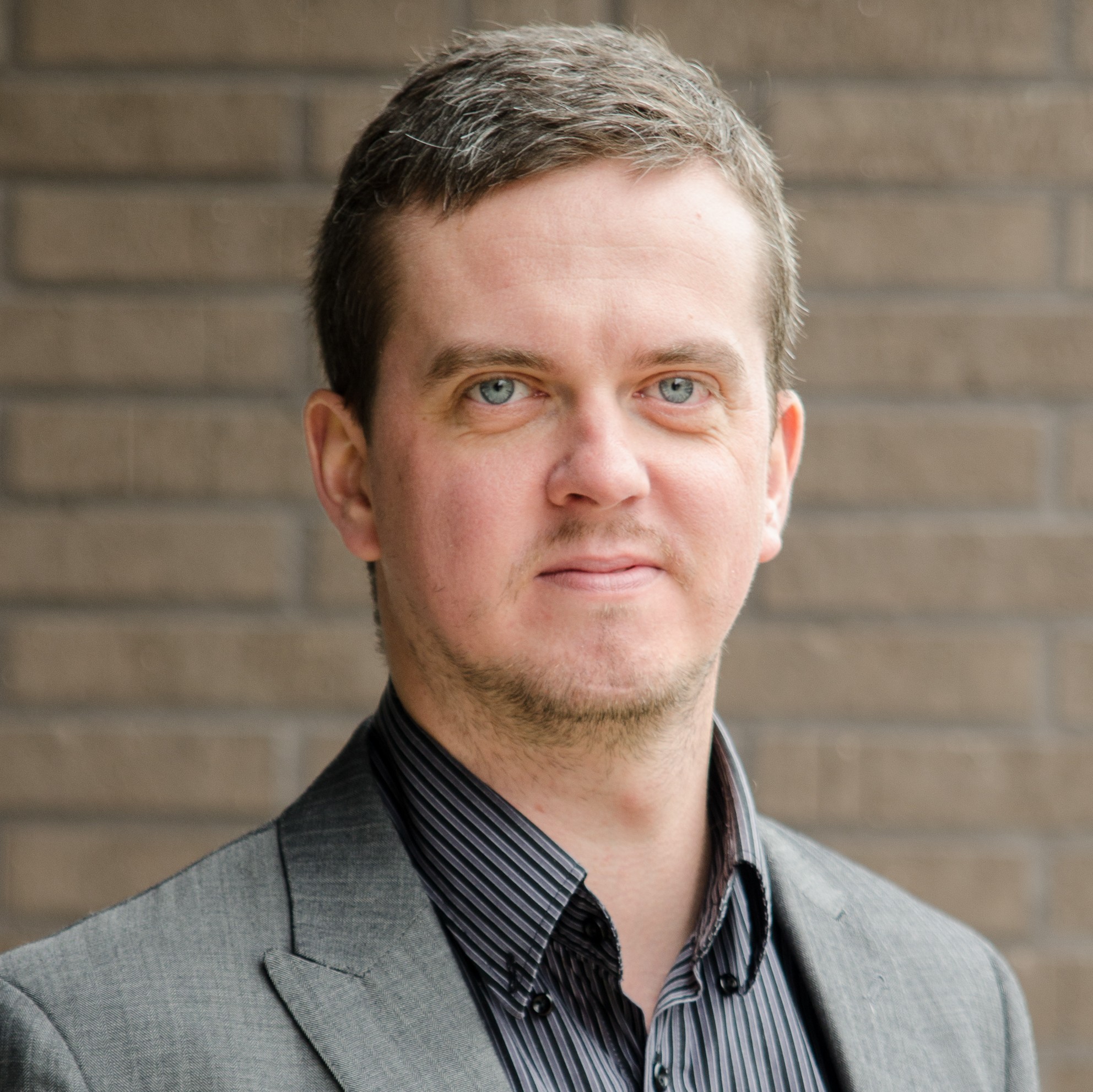}}]{Prof. Bruno Volckaert} is professor advanced software engineering and secure distributed systems at Ghent University and imec's IDLab group. His research deals with reliable and high-performance distributed software for a.o. scalable data ingestion and processing, secure software architectures and autonomous optimization of cloud/edge-based applications. He has worked on over 80 national and international research projects and is author or co-author of more than 250 peer-reviewed papers published in international journals and conference proceedings.
\end{IEEEbiography}

\EOD

\end{document}